\documentclass{aa}
\usepackage{graphicx}
\usepackage{color}
\begin{document}
\title{Very high energy gamma-ray production inside the massive binary system
Cyg~X-1/HDE~226868}

   \author{W. Bednarek$^1$ \& F. Giovannelli$^2$}

   \offprints{W. Bednarek}

   \institute{$^1$Department of Experimental Physics, University of \L \'od\'z,
              90-236 \L \'od\'z, ul. Pomorska 149/153, Poland\\
             $^2$ INAF - Istituto di Astrofisica Spaziale e Fisica Cosmica - Sezione di Roma,
Area di Ricerca CNR di Roma-2, Via Fosso del Cavaliere, 100 - I 00133 Roma, Italy\\
             \email{bednar@fizwe4.fic.uni.lodz.pl; franco.giovannelli@iasf-roma.inaf.it}
             }

   \date{Received ...., .... ; accepted ...., ....}

\abstract{ TeV $\gamma$-ray emission has been recently discovered
by Cherenkov telescopes from two microquasars, LS 5039 and LS I
+61$^{\rm o}$303. This emission is likely to be produced inside
the binary system since in both cases 
variability with the orbital period of the binary has been
discovered. In fact, such emission features have been recently 
predicted by the inverse Compton
(IC) $e^\pm$ pair cascade model. In this model, electrons
accelerated in the jet develop a cascade in the anisotropic
radiation of the massive star. The $\gamma$-ray spectra emerging
from the cascade strongly depend on the location of the observer
with respect to the orbital plane of the binary. Here we apply
this model in order to investigate the possible $\gamma$-ray
emission features from another compact massive binary of the
microquasar type, Cyg X-1. We conclude that the observational
constraints at lower energies (from MeV to GeV) suggest that the
spectrum of electrons injected in the jet is likely steeper than
in other TeV $\gamma$-ray microquasars. The cascade $\gamma$-ray
spectrum produced by electrons with such spectrum in Cyg X-1
should be below the sensitivities of the MAGIC and VERITAS class
Cherenkov telescopes. However, if the electron spectrum is
flatter, then the highest TeV $\gamma$-ray fluxes are predicted at
$\sim$7 hrs before and after the phase when the compact object is
in front of the massive star. We suggest that Cherenkov telescopes
should concentrate on these range of phases since the TeV flux can
vary by a factor of $\sim$20 with the period of the Cyg X-1 binary
system. Moreover, the model predicts clear anticorrelation of the
GeV and TeV $\gamma$-ray emission. This feature can be tested by
the future multiwavelength observations with the AGILE and GLAST
telescopes in the GeV energy range and the MAGIC and VERITAS
telescopes in the TeV energy range. 
 }

\titlerunning{gamma-ray production in Cyg X-1}

\maketitle

%

\section{Introduction}

The importance of high energy processes in X-ray binary systems
has been hypothesized for about two decades and in fact some
evidence (although not statistically significant) has been
reported in the literature (for review see e.g. Weekes 1988,
1992). However, only one of these early Cherenkov observations of
Cen X-3 has been later confirmed, and remains unquestioned up to the
present time (Chadwick et al.~2000, Atoyan et al.~2002). Some of
these well known binary systems have been also related to the
EGRET sources above 100 MeV due to directional coincidence since
they are close or inside of relatively large error boxes of some
EGRET sources: LS I +61 303 $\equiv$ 2CG135+01 $\equiv$ 2EG
J0241+6119 (Thompson et al., 1995, Kniffen et al.~1997), Cyg X-3
$\equiv$ 2EG J2033-4112 (Mori et al. 1997), Cen X-3 (Vestrand,
Sreekumar \& Mori, 1997), and LS 5039 $\equiv$ 3EG J1824-1514
(Paredes et al.~2000). However, only very recently some binary
systems belonging to the class of high mass binaries, have been
explicitly detected at high energy $\gamma$-rays by 
Cherenkov telescopes. The HESS collaboration reported the detection of TeV
$\gamma$-ray emission from the Be type binary system  PSR B1259-63,  
close to the periastron passage (Aharonian et al.~2005). Also TeV $\gamma$-rays
have been detected from LS 5039 (HESS collaboration, Aharonian et
al.~2005, Aharonian et al.~2006)  and  LS I +61 303 (MAGIC collaboration, Albert
et al.~2006). These two objects belong to the class of the so-called 
microquasars.  These objects show evidences of anisotropic outflows
usually interpreted as jet structures similar to the
relativistic jets observed in active galactic nuclei. Such jets in
microqusars might be created in the inner parts of the accretion
disks around compact objects such as a neutron or a solar mass
black hole that is accreting matter from the companion star.

In fact, microquasar features, very similar to those observed in
LS 5039 and LS I +61 303, are also observed in the well-known 
object, Cyg X-1. It has been detected by the COMPTEL
instrument up to $\sim 10$ MeV with a spectrum very close to a
power law with differential spectral index 2.6 (McConnell et
al.~2000, 2002). However, the source has not been observed at
higher energies by the EGRET instrument (see the upper limit in
McConnell et al.~ 2000). The MeV $\gamma$-ray emission is usually
interpreted in terms of two component thermal and non-thermal
comptonization models from the accretion disk (e.g. Gierli\'nski
et al.~1999), or as a result of comptonization of the stellar
photons by relativistic electrons inside the jet (Bednarek et
al.~1990, Georganopoulos et al.~2002, Romero et al.~2002). The general
picture of the production of $\gamma$-rays is that they arise as the 
result of
comptonization of stellar photons by electrons in the jet.  
This model has been
recently applied to the microquasar LS 5039 (Dermer \&
B\"ottcher~2006 and Bosch-Ramon, Romero \& Paredes~2006). However,
these models do not take into account Inverse Compton (IC) $e^\pm$
pair cascade processes.  Cascade processes become important for these binary
systems since the optical depth for electrons (which are isotropic
in the jet) depends largely on their temporary directions (for
detailed calculations of the optical depths, see Bednarek~1997,
2000 and more recently e.g. Sierpowska \& Bednarek~2005,
B\"ottcher \& Dermer~2005, Dubus~2006, Bednarek~2006a). Due to
this effect, electrons lose energy mainly in directions for which
the optical depths are the largest, but not in the direction toward
the observer (see the previously cited papers). The
calculations of the $\gamma$-ray spectra, which are produced in
the anisotropic IC $e^\pm$ pair cascade toward the observer, and
their light curves as a function of the phase of the binary
system, have been recently calculated by (Bednarek~2006a). These
calculations are based on the previous work done for other compact
binary systems such as Cyg X-3 (Bednarek~1997), and Cen X-3
(Bednarek~2000, Bednarek \& Giovannelli~1999). Recently, the
cascade model for the high energy processes inside compact massive
binaries has been successfully applied to the binary system  LS I
+61 303, predicting the phases at which the maximum TeV
$\gamma$-ray emission can be expected (Bednarek~2006a, b). Such a prediction
was confirmed by MAGIC observations (see Albert et al.~2006). As an
alternative scenario,  hadronic production of $\gamma$-rays and neutrinos
from these binaries has been also considered (e.g. Romero,
Christiansen \& Orellana~2005, and Aharonian et al.~2006).

In view of the recent discoveries of the TeV $\gamma$-ray emission
from microquasars LS~5039 and  LS~I~+61 303, we consider
the IC $e^\pm$ pair cascade model for possible high energy
processes in another binary system, Cyg X-1/HDE 226868.

\section{The binary system Cyg X-1/HDE 226868}

Cygnus X-1 has been the most extensively studied stellar mass black hole
candidate since its discovery as one of the most intense X-ray
sources in the sky (Bowyer et al.~1965). In the X-rays the source shows two
emission states (see, e.g. Priedhorsky, Terrell \& Holt~1983, Ling et
al.~1983, Liang \& Nolan~1983), spending most of the time
($\sim$90$\%$) in the so-called hard state (low level of soft X-ray
emission) and the remaining of the time in the soft state. Cyg X-1 was
also identified as a radio source by Braes \& Miley~(1971). High
resolution radio observations show a steady jet from this source
in the hard X-ray spectral state (Stirling et al.~2001) and also
a radio and optical lobe oriented along the jet direction (Gallo et
al.~2005). The jet appears when the X-ray spectrum starts
softening and its intrinsic velocity is estimated on $\ge 0.3$c
(Fender et al.~2006).

Hjellming \& Wade~(1971) identified Cyg X-1 with the optical star
HDE 226868, which is an O9.7 Iab type massive star inside a 
binary system with a period of 5.6 days (Webster \& Murdin~1972,
Bolton~1972). The main parameters of this binary system have been
recently confirmed by Zi\'o\l kowski~(2005): the mass of the optical
companion M$_\star = 40\pm 10$ M$_\odot$; its radius R$_\star = 22
\pm 3$ R$_\odot$; an effective temperature of T$_\star = 30,000 \pm
2,000$ K for a distance to the system D = 1.8--2.35 Kpc; and a mass of
the compact object M$_{\rm x} = 16 \pm 3$ M$_\odot$ which is in an
almost circular orbit $e=0$; with an inclination angle of the
binary $i = 33^\circ \pm 5^\circ$. In this paper we adopt the
values  for the binary system shown in Table~1.

\begin{table}
\caption{Parameters of the binary system and the jet.}             
\label{tab2}      
\begin{tabular}{c c}     
\hline\hline       
            Parameter      &  Value \\
            \hline
            \noalign{\smallskip}
            $R_\star$: stellar radius      & 20 $R_\odot$             \\
            $T_\star$: surface temperature  & $3\times 10^4$ K            \\
            $M_\star$: stellar mass           &   $40\pm 10$ M$_\odot$  \\    
            $M_{\rm x}$: companion mass           &   $16\pm 3$ M$_\odot$  \\    
           a: orbital radius          & $2.15R_\star$  \\
            e: eccentricity                                & 0              \\
            i: inclination                              & $30^{\rm o}$     \\
            \hline
            $r_{\rm in}$: disk inner radius, radius of the jet   &   $10^7$ cm \\
            $\alpha$: jet opening angle                          &   0.1 rad  \\
            $z_{\rm min}$: inner boundary of the {\it active} jet  &  $0.1R_\star$  \\
            $z_{\rm max}$: outer boundary of the {\it active} jet  &  $10R_\star$  \\
            $\eta$: magnetization parameter of the jet    & 1.0-0.01 \\
	    $B_{\rm d}$: magnetic field at the base of the jet (for $\eta=1$) & $10^5$ G  \\
	    $\xi$: acceleration efficiency of electrons       & 0.5-0.01             \\
\hline                  
\end{tabular}
\end{table}

\section{A microquasar in the compact massive binary}

We consider a generally accepted scenario for microquasars,
in which electrons are accelerated continuously along the jet. The
jet propagates from the compact object along an axis perpendicular to the
plane of the binary system. The details of such a general scenario
have already been  described in Bednarek~(2006b). Here we repeat
the basic points of the scenario for the sake of completeness. The shock
acceleration mechanism is usually accepted as being responsible for the
acceleration of the electrons. It is assumed that electrons are
injected with the power law spectrum extending up to the maximum
energies determined by the balance between the acceleration rate
and the most efficient energy loss rate. In the case of
microquasars, the synchrotron process (determined by the
magnetic field strength locally in the jet) or the inverse Compton
process (determined by the radiation field of the massive star)
usually provide the most efficient cooling for the electrons.

The magnetic
field strength inside the jet can be estimated on the basis of the
equipartition arguments concerning the relation between the accretion disk
luminosity and the kinetic power of the jet, i.e. the so called
jet-accretion-disk symbiosis model developed by Falcke et
al.~(1995), Falcke et al.~(2004).  In fact, in the case of Cyg X-1, 
the X-ray luminosity (probably produced in the accretion disk) and 
the jet kinetic power are estimated to be 
of the same order  (Gallo et al. 2005). Note, however, that this is not 
always the case since some other binaries are clearly underluminous,
e.g. LS 5039 and LS I +61 303. Moreover, it is commonly
expected, on the basis of the equipartition arguments, that the
energy density of the magnetic field inside the jet should be
related to the kinetic power of the jet (see e.g. Bosch Ramon et
al.~2005). Therefore, we accept the hypothesis 
that the magnetic field energy
density in the inner part of the accretion disk (at the base of
the jet) should be related to the energy density of the disk
radiation (see also e.g. Bednarek~2005). On the basis of these
arguments we arrive to the conclusion that the magnetic field in
the jet can be as large as $B_{\rm d}\sim 10^5$ G (e.g.
Bednarek~2006b). Assuming a simple conical structure for the jet
(fixed opening angle $\alpha$), the magnetic field in the jet
should drop with the distance from its base according to
\begin{eqnarray}
B(x) = B_{\rm d}\eta/(1 +\alpha z)\approx B_{\rm d}\eta/(\alpha z)~~~{\rm
for~~~z\gg 1/\alpha},
\label{eq1}
\end{eqnarray}
where $\alpha$ is assumed of the order of $0.1$ rad, $\eta$ is the
ratio of the energy density of the magnetic field to the energy
density of the disk radiation at the base of the jet 
(magnetization parameter of the jet). 
This parameter describes how far the magnetic field is from the
equipartition with respect to the kinetic energy density of the jet
plasma. $z$ is the distance along the jet from its base expressed
in units of the disk inner radius $r_{\rm in}$, where $r_{\rm in}$
is of the order of $10^7$ cm for the compact object with the mass
of the order of 10 M$_\odot$.

\begin{figure*}
\vskip 5.3truecm 
\includegraphics{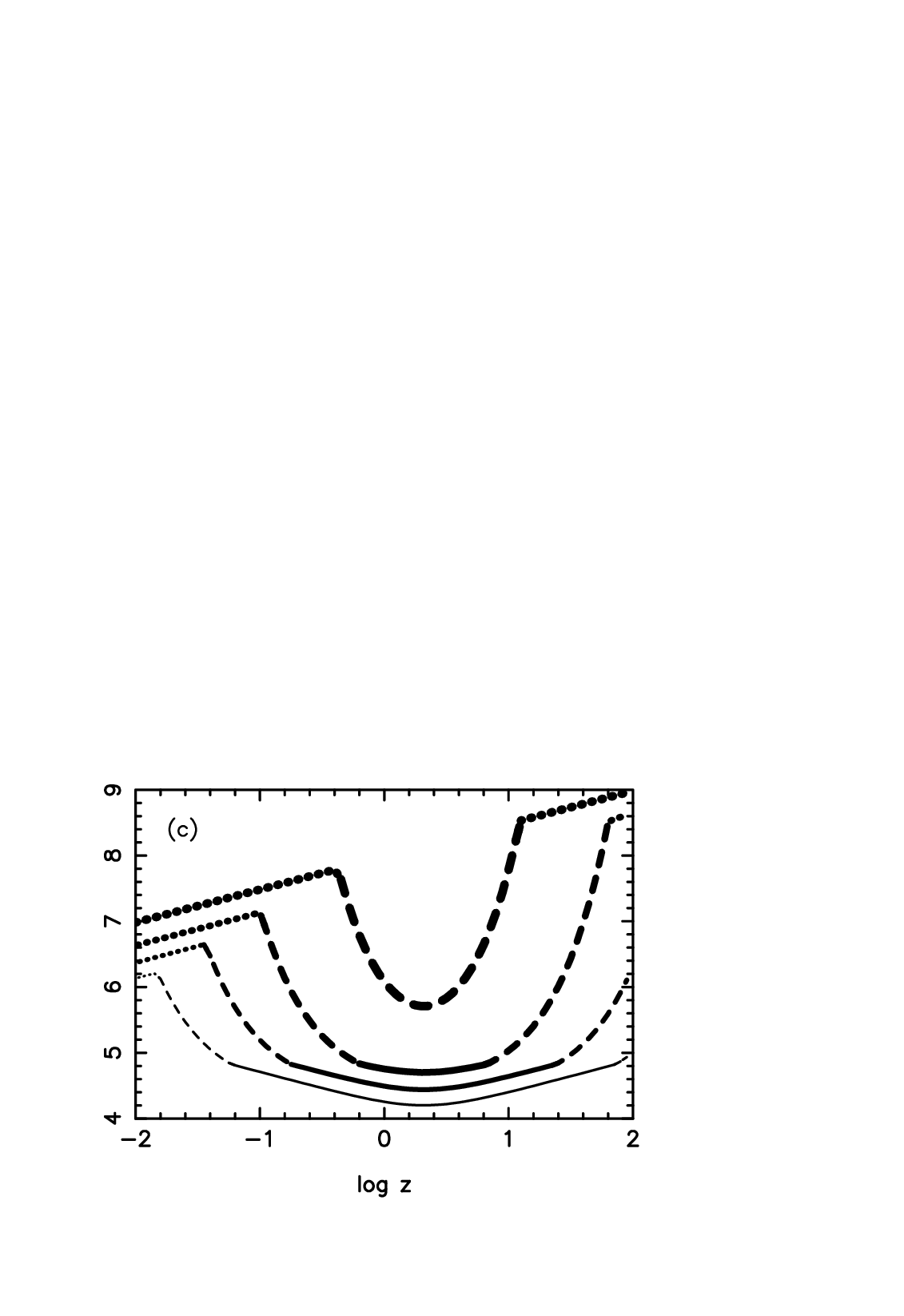} 
\includegraphics{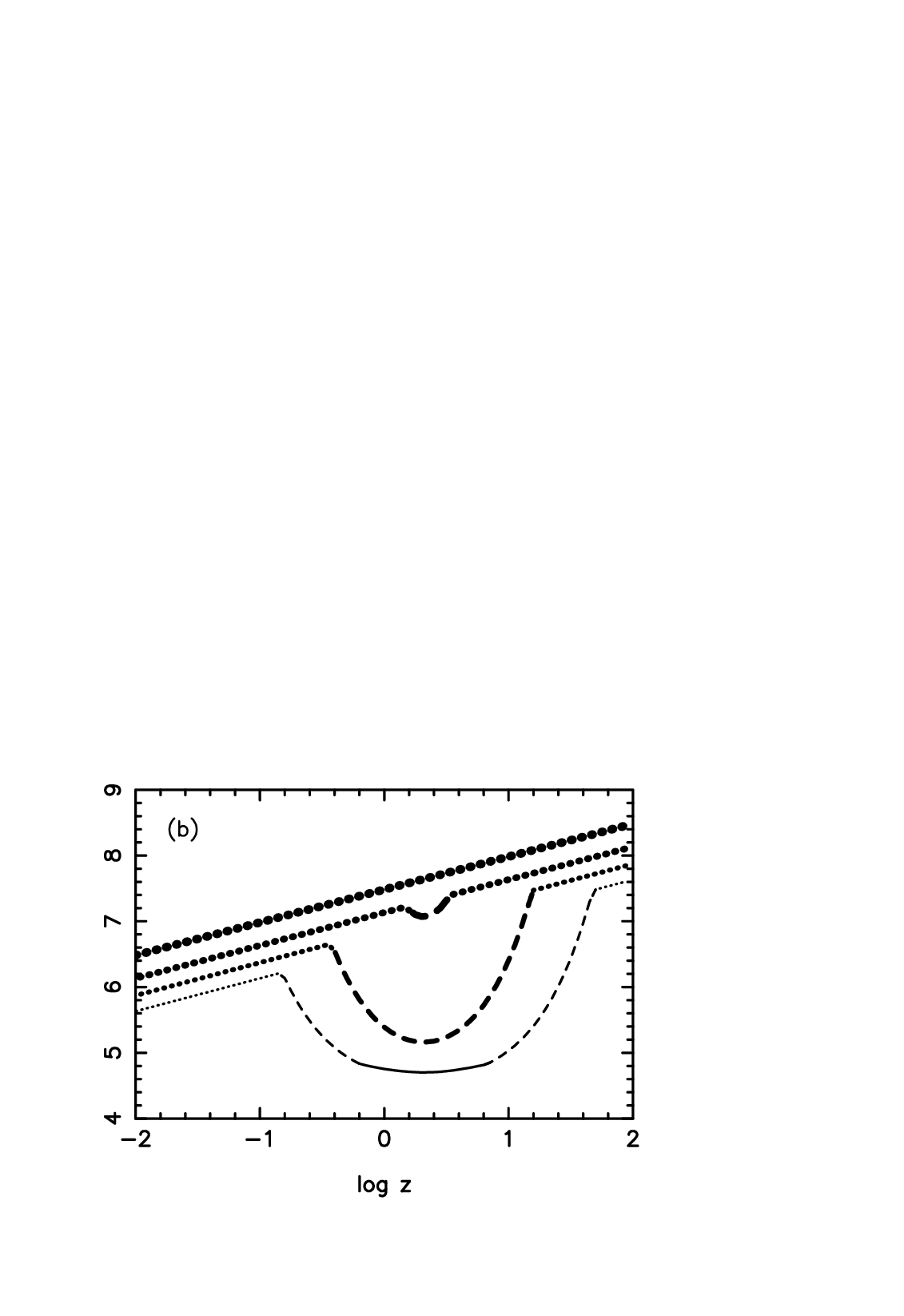}
\includegraphics{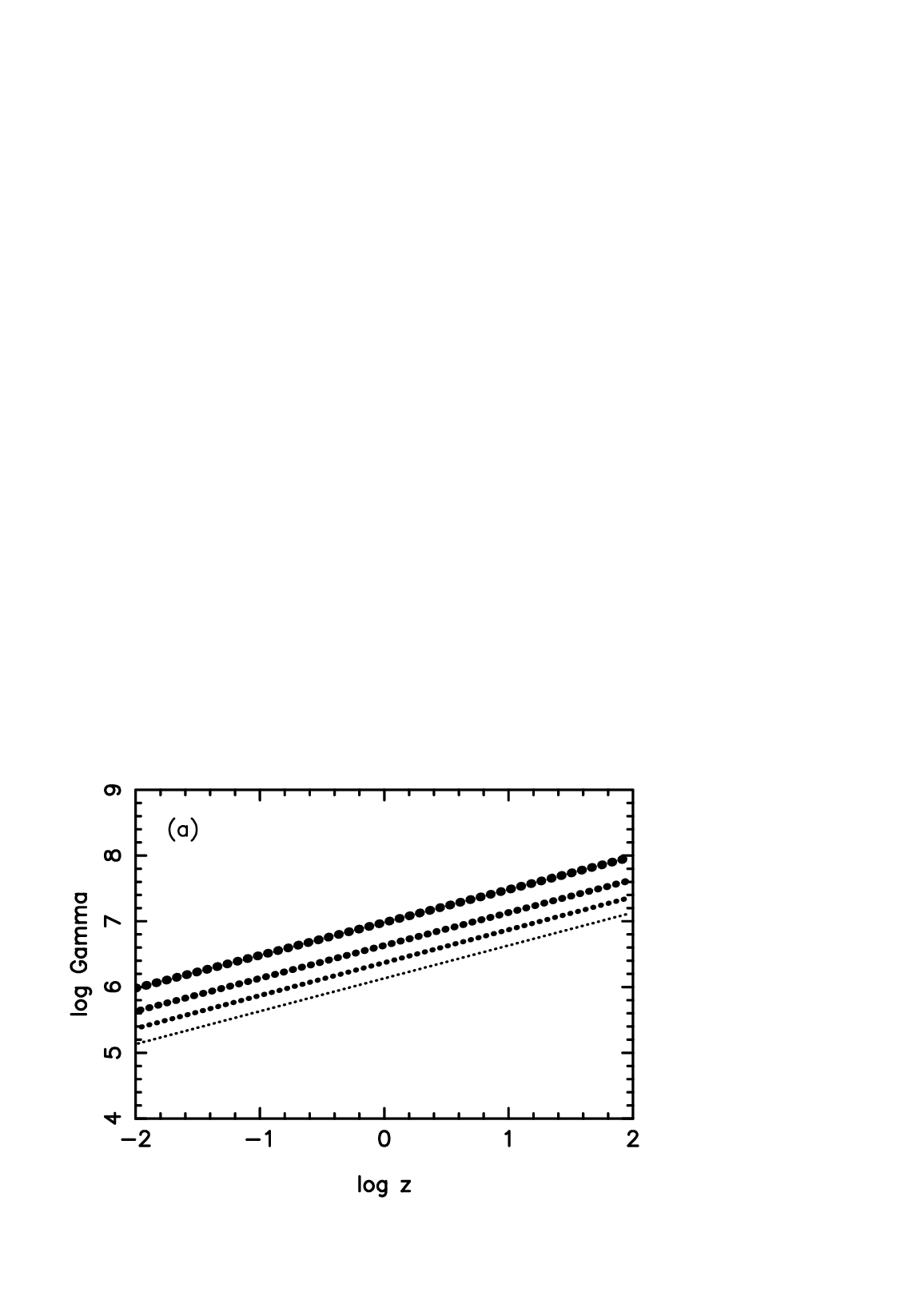} \caption{The maximum Lorentz factors of electrons as a
function of the distance, $z$ from the base of the jet (measured
in units of the stellar radius $R_\star$) for different
acceleration efficiencies: $\xi =0.5$ (from the upper, thickest
curve), 0.1, 0.03, and 0.01 (to the bottom, finest curve) and two
ratios of the magnetic field to disk radiation energy densities at
the base of the jet equal to $\eta=1$ (a), 0.1 (b), and 0.01 (c).
The Maximum Lorentz factors are obtained from the comparison of
the acceleration rate with the energy loss rate on synchrotron
process (dotted lines), inverse Compton process on scattering of
stellar radiation from the massive star HDE 226868 in the Thomson
regime (full curve) and the Klein-Nishina regime (dashed curves).
The base of the jet is at the distance of  2.15$R_\star$ from the
center of the massive star.} 
\label{fig1}
\end{figure*}

Having defined the magnetic field strength at different parts of
the jet, we can estimate the acceleration rate of the electrons
(with energy $E$ and the Lorentz factors $\gamma$),
\begin{eqnarray}
\dot{P}_{\rm acc}(\gamma) = \xi c E/r_{\rm L}\approx 10^{13}\xi B~~~{\rm eV~s^{-1},}
\label{eq2}
\end{eqnarray}
\noindent where $\xi$ is the acceleration efficiency, $r_{\rm L} =
E/eB$ is their Larmor radius, $B$ is the magnetic field at the
acceleration site (in Gauss), $e$ is the electron charge, and $c$
is the velocity of light. In Eq. 1, a fundamental role is played by
the acceleration efficiency of particles $\xi$, but this is in fact
unknown. There are some observational suggestions that $\xi$
should be not far from unity in the case of relativistic shocks,
e.g. presence of relativistic leptons with energies $\sim 10^{15}$
eV inside the Crab Nebula. 
On the other hand, the theory of shock acceleration processes
estimates $\xi$ as $\sim 0.1(v/c)^2$, i.e. $> 0.01$ for $v>0.3c$
(Malkov \& Drury~2001). Therefore, in this paper we consider the
values of the acceleration efficiency of electrons in the jet in
the range $\xi = 0.5-0.01$. The basic parameters which describe 
the jet are shown in Table~1.

As we noted above, during the acceleration process, electrons lose
energy mainly by the synchrotron and the inverse Compton (IC)
processes. We are interested in situations in which electrons
can be accelerated to TeV energies. This is not easy to obtain
relatively close to the base of the jet where the magnetic field
is large and the synchrotron losses very strong. Therefore, since
we are interested in efficient injection of high energy IC
$\gamma$-rays by electrons in the jet, it makes sense to consider
distances from the accretion disk for which the IC energy losses
on the massive star radiation are at least comparable to 
synchrotron losses. In this cascade model we will concentrate on
the parameter range adequate for the production of TeV $\gamma$-rays.

Let us estimate the maximum energies of electrons in the shock
acceleration process allowed by the energy losses. 
In these estimations we concentrate on the parts of the jet which are above
$z_{\rm min} = 0.1R_\odot = 1.4\times 10^{11}$ cm, since very close to the base of the jet the 
radiation field from the accretion disk (and disk corona) may dominate over 
the massive star radiation field (total luminosity of the disk is estimated in 
several times $10^{37}$ erg s$^{-1}$, see e.g.  Poutanen, Krolik \& Ryde~1997). 
At $z_{\rm min}$, the disk radiation field is diluted by the factor 
of $\sim (\beta z_{\rm min}/r_{\rm in})^2\approx 10^{-6}-10^{-4}$, where
$r_{\rm in}\sim 10^7$ cm is the inner radius of the disk around the $10M_\odot$ black hole,
and the parameter $\beta = r_{\rm in}/r_{\rm cor}\approx 0.01-0.1$ defines the extend of the disk corona assumed in the range of $r_{\rm cor}\approx 10-100r_{\rm in}$.
Note that the massive star luminosity in Cyg X-1 is clearly larger than the disk luminosity.
Therefore, disk radiation can become important only in a relatively small part of the inner jet.

The energy loss rate by the synchrotron process is given by,
\begin{eqnarray}
{\dot P}_{\rm syn}(\gamma) = {{4}\over{3}}\pi \sigma_{\rm T} c U_{\rm B}\gamma^2
\approx 2.7\times 10^{-3}B^2\gamma^2~~~{\rm eV~s^{-1},}
\label{eq3}
\end{eqnarray}
\noindent where $U_{\rm B} = B^2/8\pi$ is the energy density of
the magnetic field, $\gamma$ is the Lorentz factor of the
electrons, and $\sigma_{\rm T}$ is the Thomson cross section.
Electrons lose energy on the IC process in the Thomson (T) regime,
if their Lorentz factors are,
\begin{eqnarray}
\gamma\ll \gamma_{\rm T/kN} = m c^2/3k_{\rm B}T\approx 2\times 10^5/T_4,
\label{eq3a}
\end{eqnarray}
(where $T = 10^4T_4$ K is the surface temperature of the massive
star which defines the black body spectrum, and $k_{\rm B}$ is the
Boltzman constant), and in the Klein-Nishina (KN) regime, for the
Lorentz factors $\gamma\gg \gamma_{\rm T/KN}$. The energy loss
rate in the Thomson regime is given by
\begin{eqnarray}
{\dot P}_{\rm IC}^{\rm T}(\gamma) = {{4}\over{3}}\pi \sigma_{\rm T} c
U_{\rm rad}\gamma^2 \approx 3.8T_4^4\gamma^2/r^2~~~{\rm eV~s^{-1},}
\label{eq4}
\end{eqnarray}
\noindent where $U_{\rm rad} = 4.7\times 10^{13}T_4^4/r^2$ eV
cm$^{-3}$, and $r$ is the distance to the center of the massive
star in units of its radius $R_\star$. The energy loss rate in the
KN regime depends only logarithmically on the Lorentz factor of
the electrons. We approximate these losses by (see e.g. Blumenthal
\& Gould 1970),
\begin{eqnarray}
{\dot P}_{\rm IC}^{\rm KN}(\gamma)\approx
{\dot P}_{\rm IC}^{\rm T}(\gamma_{\rm T/KN})\ln{(4k_{\rm B}T\gamma/mc^2-2)}.
\label{eq5}
\end{eqnarray}
\noindent The maximum energies of the accelerated electrons are
determined by the balance between the acceleration mechanism and
by the most efficient mechanism for the energy losses. The
comparison of Eqs.~\ref{eq2} and~\ref{eq3} gives the maximum
allowed Lorentz factors of the electrons due to the synchrotron
energy losses,
\begin{eqnarray}
\gamma_{\rm syn}\approx 6\times 10^7(\xi/B)^{1/2}.
\label{eq6}
\end{eqnarray}
The maximum energies allowed by the IC process in the Thomson
regime are (from Eqs.~\ref{eq2} and~\ref{eq4}),
\begin{eqnarray}
\gamma_{\rm IC}^{\rm T}\approx 1.6\times 10^6(\xi B)^{1/2}r/T_4^2,
\label{eq7}
\end{eqnarray}
and in the KN regime are (from Eqs.~\ref{eq2} and~\ref{eq5}),
\begin{eqnarray}
\gamma_{\rm IC}^{\rm KN}\approx 1.5\times 10^5(2+e^{(71\xi B r^2/T_4^2)})/T_4.
\label{eq8}
\end{eqnarray}
Note that the limit given by Eq.~\ref{eq7} is only valid provided
that $\gamma_{\rm IC}^{\rm T}<\gamma_{\rm T/KN}$, i.e when the
parameters of the jet and the massive star fulfill the following
condition, $T_4/r > 8(\xi B)^{0.5}$ (from Eqs.~\ref{eq3a}
and~\ref{eq7}). In any other case, the limit given by
Eq.~\ref{eq8} has to be taken.

The Larmor radii of electrons accelerated to the maximum Lorentz factors 
allowed by the radiation constraints  has to be smaller than expected  
from the constraints imposed by the size of the jet. The Larmor radius  
of electrons (defined below Eq.~2) is thus lower than the size of the jet,
$r_{\rm L}<(1 + \alpha z)r_{\rm in}$ (see above), provided that
$\gamma_{\rm size}< eB_{\rm d}\eta/m_{\rm e}$. For the considered parameters and 
the applied model for the magnetic field distribution along the jet  (see Eq.~1),  
$\gamma_{\rm size}< 6\times 10^8$. Note that this limit is independent of the 
distance from the base of the jet. It does not impose additional constraints on the 
values obtained above for distances from the base of the jet which are of interest in this model.

The maximum Lorentz factors to which electrons can be accelerated
at a specific part of the jet are determined by the acceleration
efficiency and the energy losses through synchrotron and IC
processes. These energy loss processes depend on the conditions in
the jet (i.e., magnetic field strength and acceleration efficiency), on the
parameters of the massive star (its radius and surface temperature),
and on the distance of the acceleration region from the massive
star.  In Fig.~\ref{fig1}, we plot the maximum Lorentz factors of
electrons as a function of the distance from the base of the jet
(given by either Eq.~\ref{eq6}, or Eq.~\ref{eq7}, or
Eq.~\ref{eq8}), for selected values of the acceleration
efficiency, $\xi$, and the ratio, $\eta$. 
The maximum energies of the accelerated electrons
increase with the distance from the base of the jet as $\propto
z^{1/2}$, when the acceleration process is saturated by
synchrotron energy losses (see also Bosch Ramon et al.~2006).
However, in the regions of the jet where the magnetic field is
relatively weak (farther from the base of the jet but still close
to the massive star), the acceleration process is saturated by IC
energy losses. Then, the maximum Lorentz factors of the electrons
are significantly lower (note the broad parabolic dips in Fig. 1,
especially evident during the periastron passage of the compact
object).

\begin{figure*}
\vskip 5.3truecm 
\includegraphics{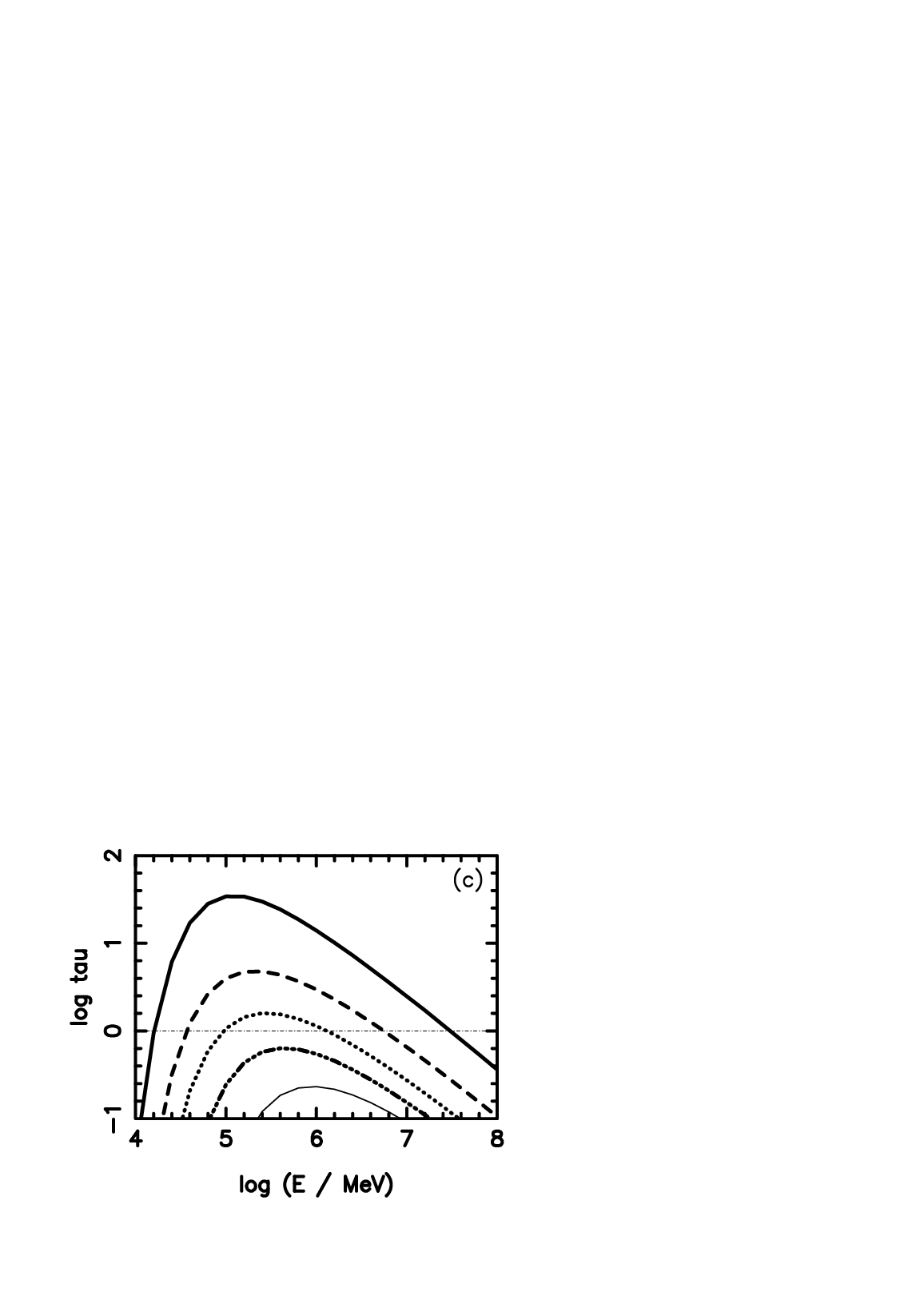} 
\includegraphics{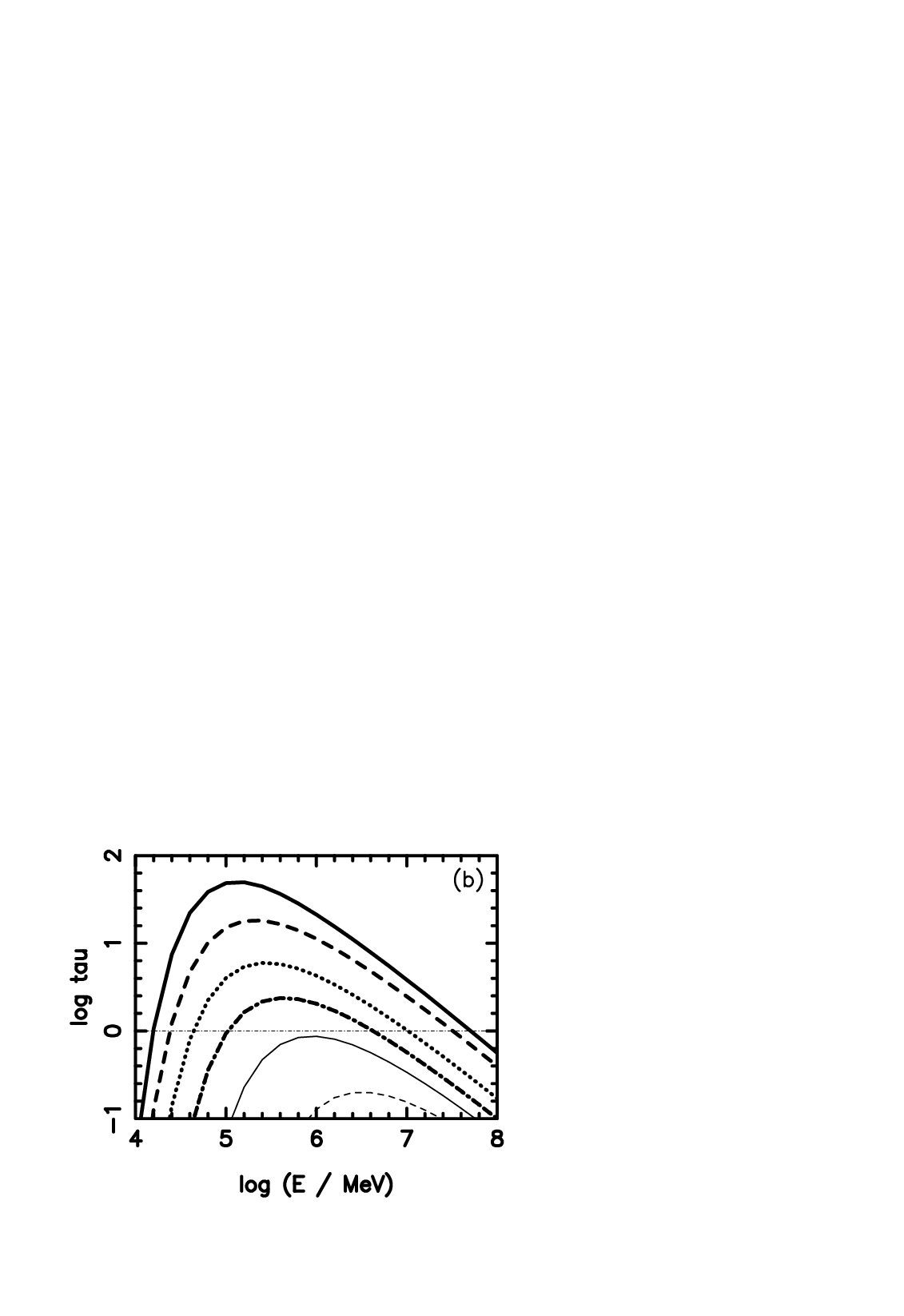}
\includegraphics{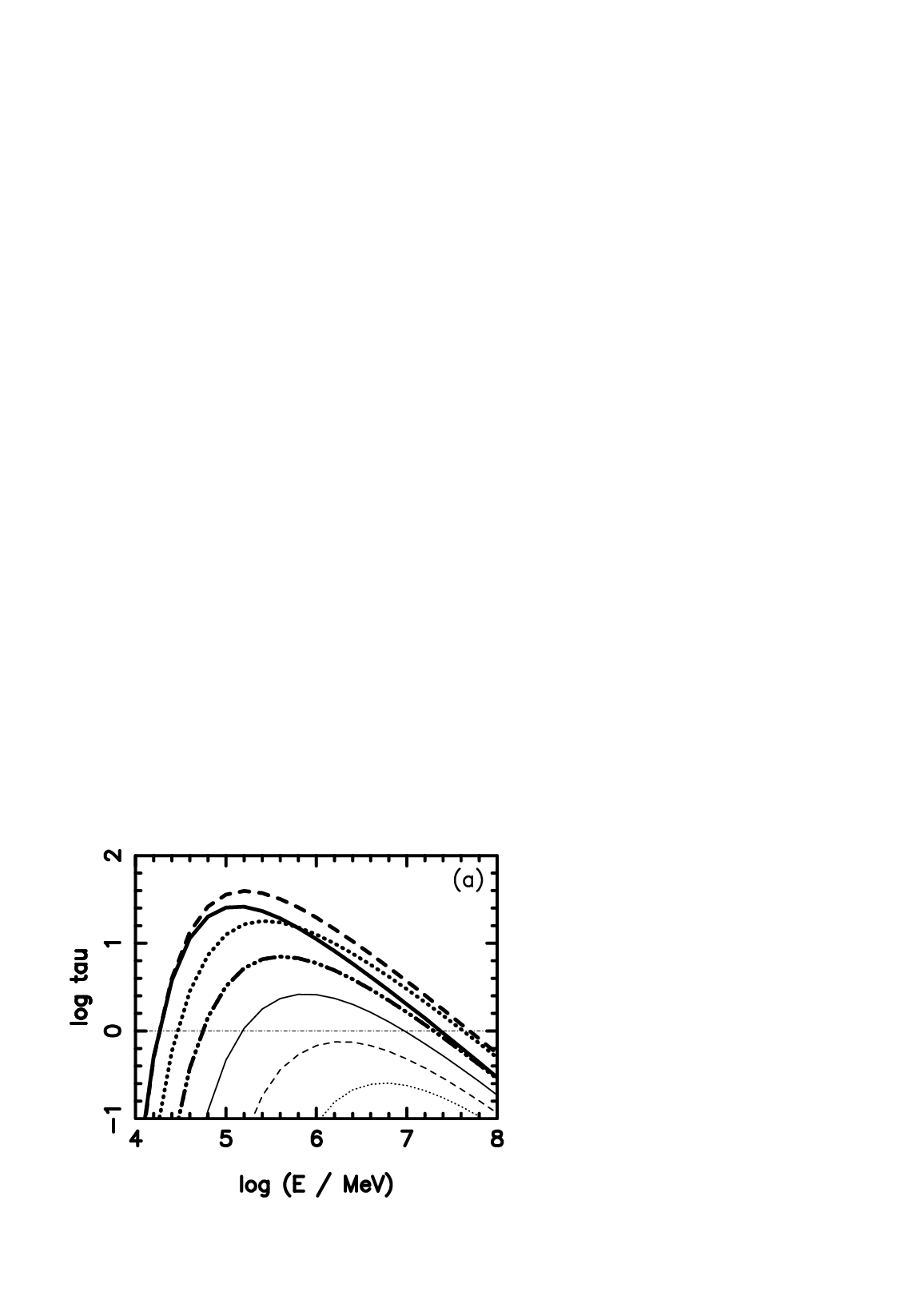} 
\caption{(a) The optical depths for $\gamma$-rays (as a
function of their energy) on $e^\pm$ pair production in collisions
with stellar photons calculated for $\gamma$-rays propagating from
a specific distance from the massive star up to the infinity.
$\gamma$-rays are injected at three different distances from the
massive star  HDE 226868, $z = 2R_\star$ (a), $z = 5R_\star$ (b),
and $z = 20R_\star$ (c), and for selected angles measured from the
direction defined by the centers of the stars,  $\alpha = 0^{\rm
o}$ (i.e. toward the center of the massive star the optical depth
is calculated only to the surface of the star, thick full curve),
$30^{\rm o}$ (thick dashed), $60^{\rm o}$ (thick dotted), $\alpha
= 90^{\rm o}$ (thick dot-dashed), $120^{\rm o}$ (thin full),
$150^{\rm o}$ (thin dashed),  and $180^{\rm o}$ (thin dotted).}
\label{fig2}
\end{figure*}

In a previous work (Bednarek~2006b), we have shown that
electrons accelerated in the jet have to cool locally, i.e. close
the the acceleration site, due to the very large energy losses,
provided that they are injected within $\sim 100R_\star$ from the
base of the jet. At larger distances accelerated electrons are
advected outside the binary system (along the jet) and can
produce high energy radiation through other processes. However
in such a case, we do not expect any clear variability for the
$\gamma$-ray emission with the period of the binary system.  
Therefore, we do not consider that possibility in this paper.

Following our previous work, we also assume that electrons are
injected with the power law spectrum described by the single
spectral index independent on the distance from the base of the
jet $z$ (typical values from the range 2--2.6 are considered).
However, the injection rate of electrons, i.e. the power
transfered from the magnetized plasma to relativistic electrons
$L_{\rm e}$, can differ, depending on a range of possible processes
occurring in the jet. Two different likely scenarios are
considered:

\begin{enumerate}

\item constant injection of electrons along the
jet from its base at $z_{\rm min}$: $L_{\rm e} = const.$,

\item injection rate drops along the jet according
to  $L_{\rm e}(z)\propto z^{-2}$.

\end{enumerate}

\subsection{Propagation of $\gamma$-rays}

Since the Cyg X-1 binary system is very compact (with a 
separation of the
stars of only $a = 2.15R_\star$), $\gamma$-rays produced in the part
of the jet which is closer than a few tens of R$_\star$ from the
massive star can be efficiently absorbed by collisions with the
thermal radiation from the massive star. Following our previous
works (Bednarek 1997, 2000, Sierpowska \& Bednarek~2005), we
calculate the optical depths for $\gamma$-rays as a function of
the energy in the most general case, i.e. for an arbitrary distance
from the massive star and direction of their propagation (for an angle
measured from the direction defined by the centers of the stars). The
results are shown in Fig.~\ref{fig2} for an injection distance
of $\gamma$-rays from the center of the massive star at $2R_\star$
(a), $5R_\star$ (b), and   $20R_\star$  (c). It is clear that the
optical depths are significantly above unity at specific
directions even if the injection distance from the massive star is
quite large. Since the cross sections for the IC process in the KN
regime and for the $\gamma$+$\gamma$ collisions
($\gamma$+$\gamma\rightarrow e^\pm$) are comparable, it is
expected that relativistic electrons injected isotropically inside
the jet lose energy mainly in directions where the optical depths
are the largest. They inject primary $\gamma$-rays mainly in
directions of the largest optical depths, i.e. toward the source
of soft radiation. These primary $\gamma$-rays can develop IC
$e^\pm$ pair cascades in the anisotropic radiation field of the
massive star. A part of these cascade $\gamma$-rays escape toward
the observer located at specific directions with respect to the
binary system. We follow such a complicated cascade process applying
the Monte Carlo cascade code described in detail in our previous
works (e.g. Bednarek~2000, 2006a). The code includes also the
effects connected with synchrotron cooling of the electrons
injected into the magnetic field of the jet. Therefore, we are
also able to calculate the simultaneous synchrotron photon spectra
produced by electrons accelerated in the jet. 
$\gamma$-ray photons escaping from the binary system are sorted in
specific ranges of solid angles on the hemisphere and grouped at
specific ranges of energies. Therefore, as a final product, we
obtain the $\gamma$-ray spectra produced in such anisotropic
cascades at specific directions on the hemisphere.

We note that in  parts of the jet relatively close to the accretion disk 
(this part of the jet is not considered in the present paper), $\gamma$-rays can
also be absorbed in the disk radiation field (see, e.g., the calculations of the optical 
depths by e.g. Herterich~1974, Protheroe \& Stanev~1987, 
Carraminana~1992, Bednarek~1993). The disk radiation can be also upscattered 
to the $\gamma$-ray energies by electrons provided that they are accelerated 
very close to the base of the jet as  considered by e.g.  
Aharonian et al. (1985) or Romero et al.~(2002).

\section{Cascade gamma-ray spectra}
\begin{figure*}
\vskip 9.5truecm 
\includegraphics{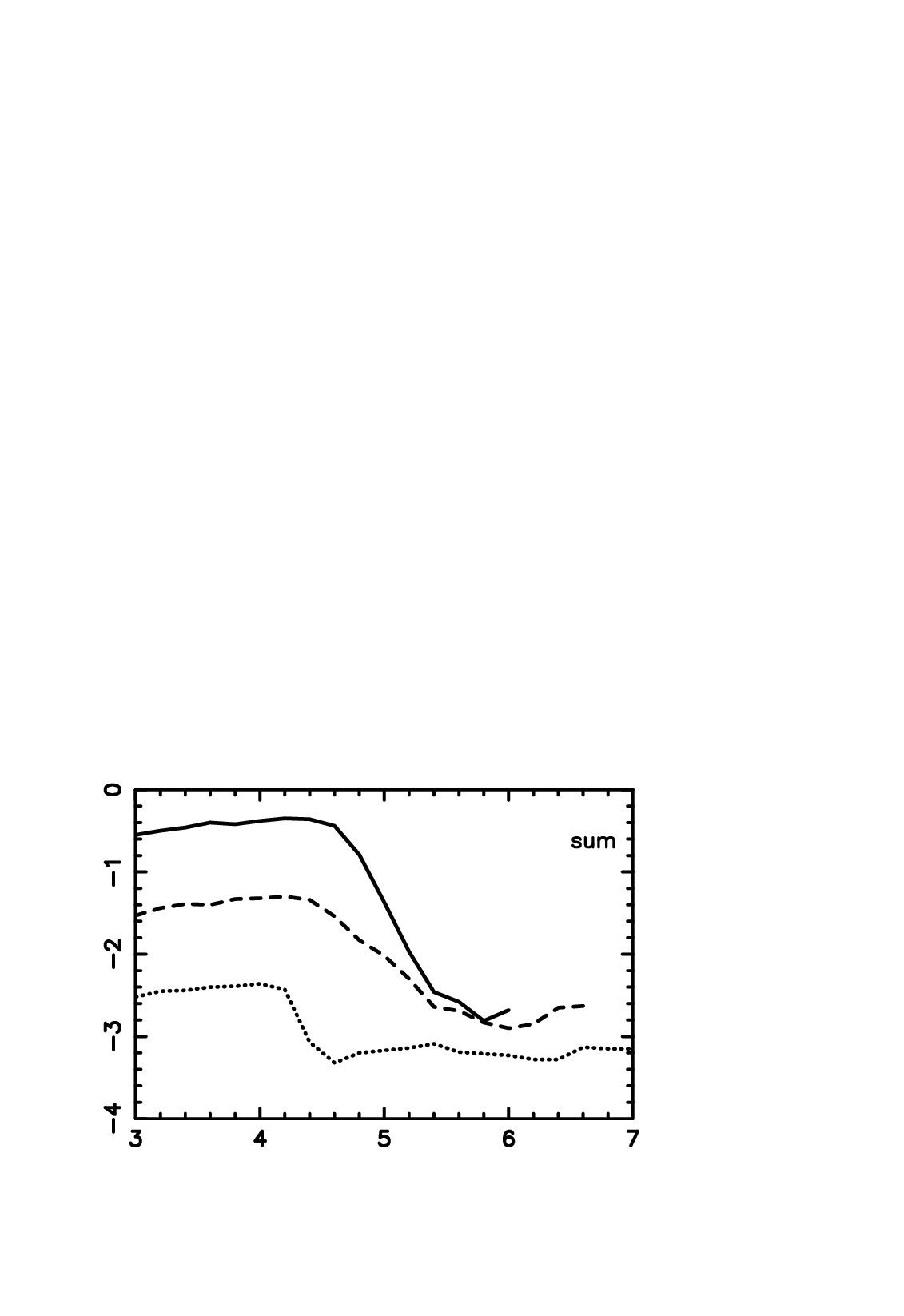} 
\includegraphics{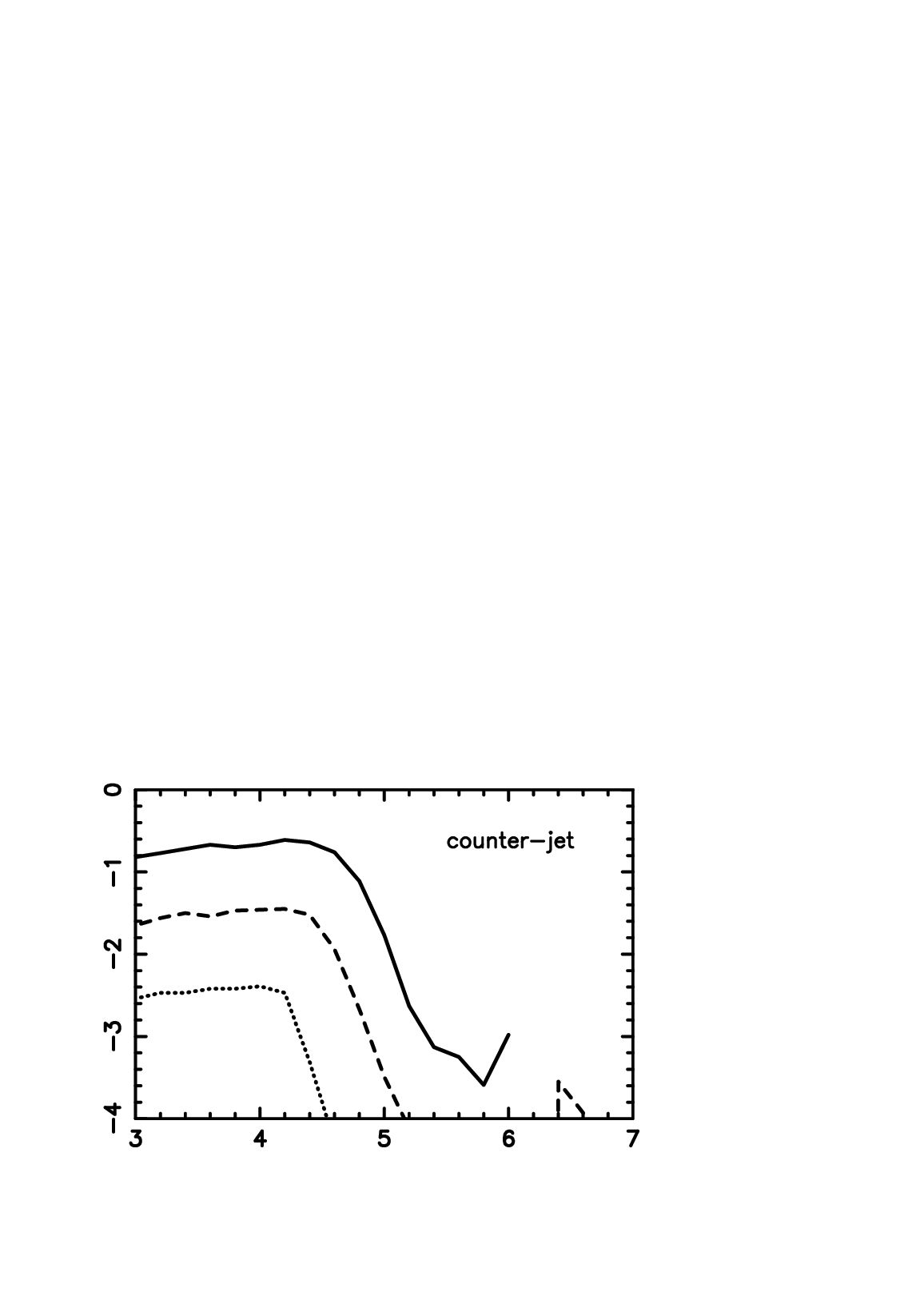}
\includegraphics{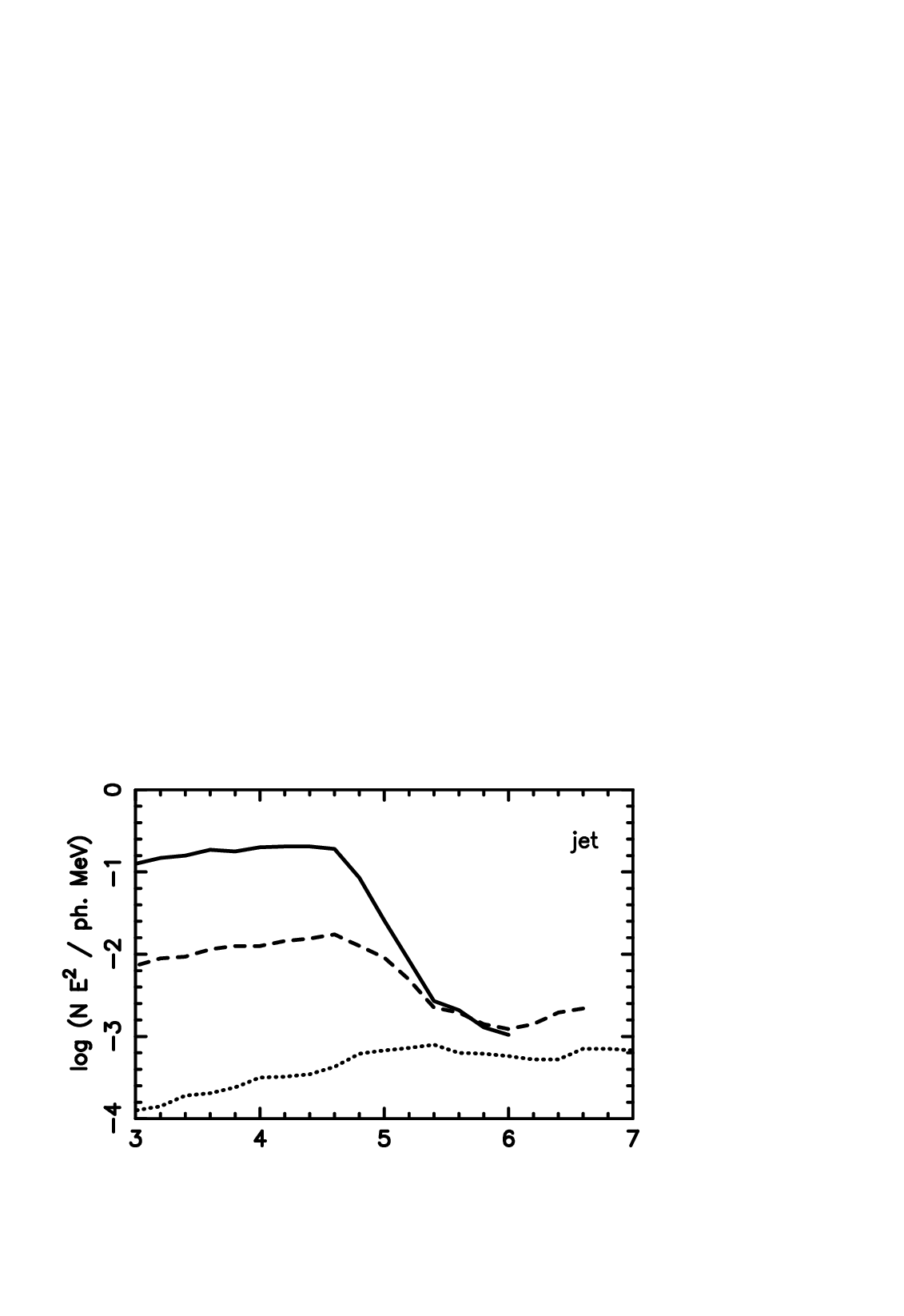} 
\includegraphics{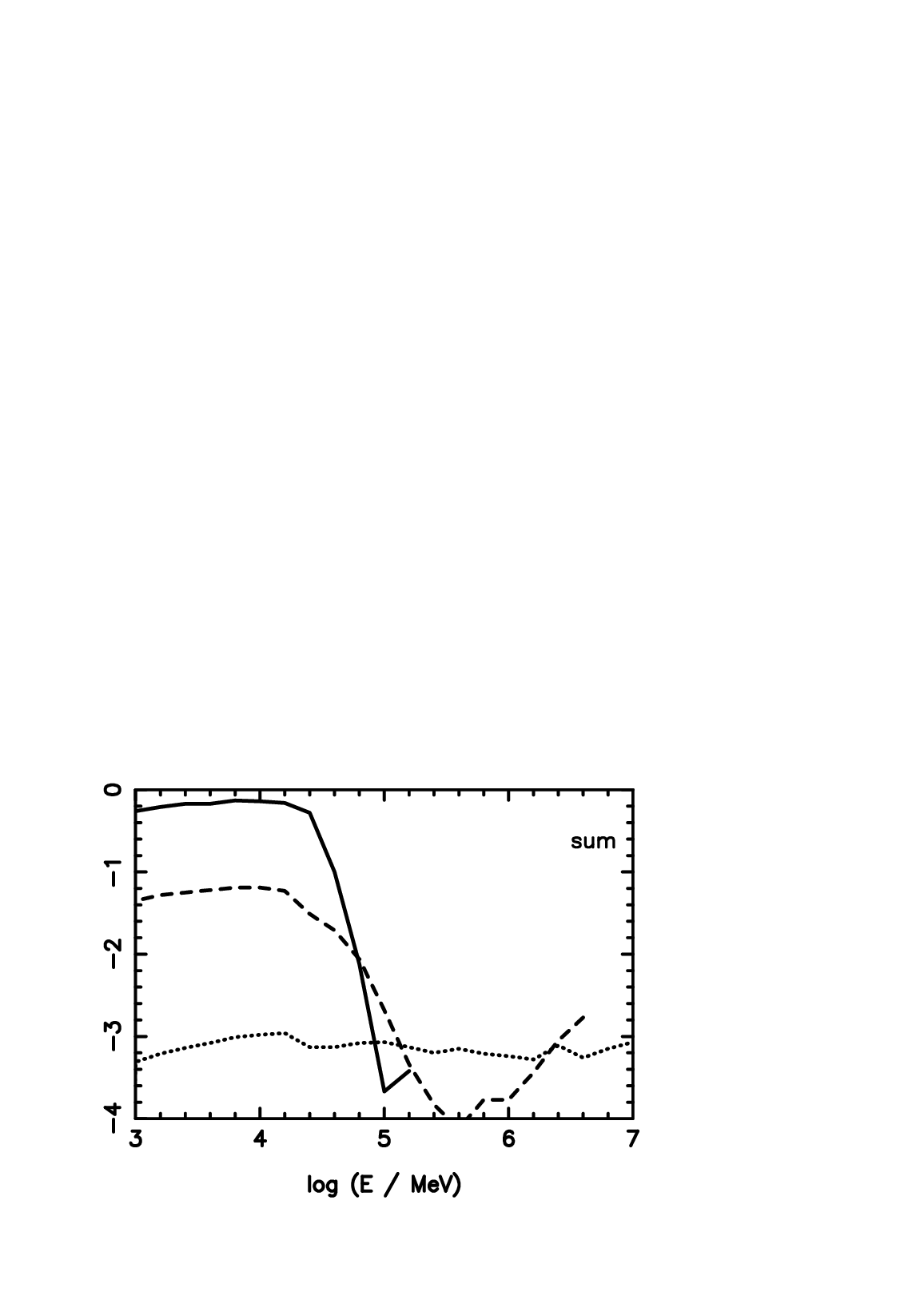} 
\includegraphics{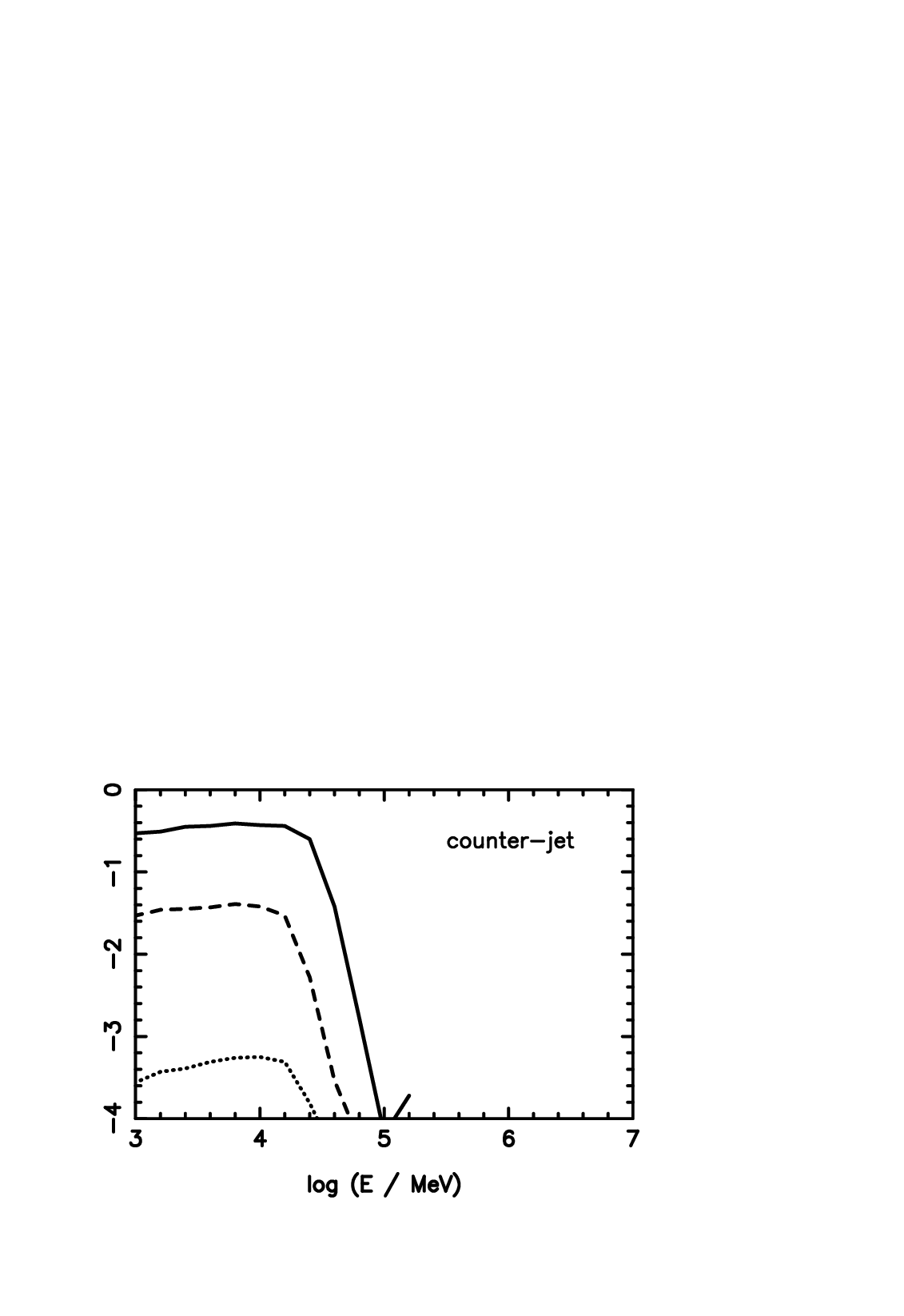} 
\includegraphics{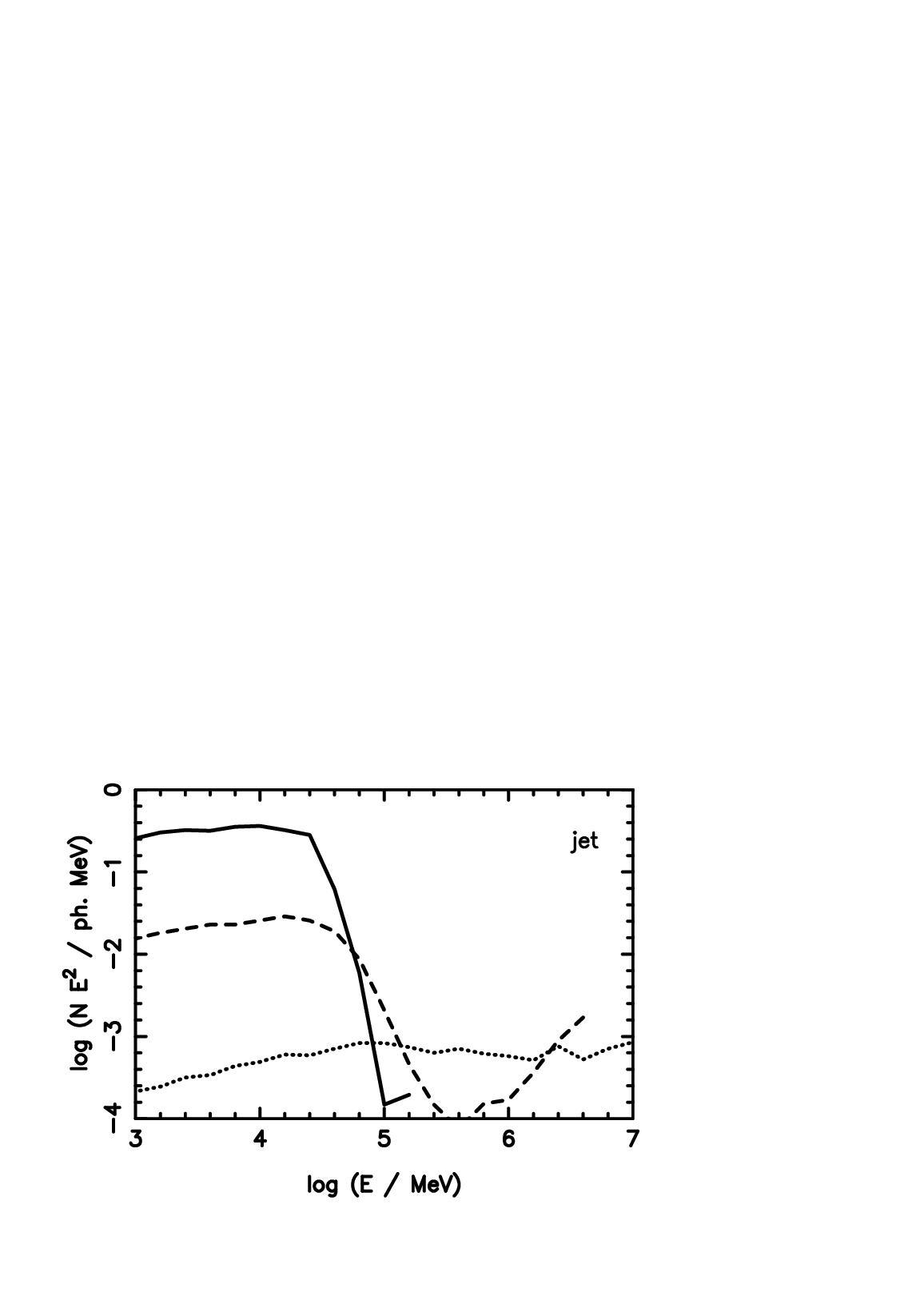} 
\caption{Cascade
$\gamma$-ray spectra produced by electrons injected at different
distances from the base of the jet $z = 0.1r_\star$ (full curve),
1$R_\star$ (dashed), and $10r_\star$ (dotted) for two locations of
the compact object on its orbit around the massive star, in front
of the massive star (upper figures) and behind the massive star
(bottom figures). The $\gamma$-ray spectra expected from the jet
moving inside the hemisphere containing the observer, who is
located at the inclination angle of $i = 30^{\rm o}$, are shown on
the  left figures, from the counter-jet (middle figures), and the
sum of the jet + counter-jet spectra (right figures). The
acceleration of electrons in the jet is defined by the parameters
$\xi = 0.1$ and $\eta = 0.1$ and the electrons are injected with a
power which drops along the jet according to $L_{\rm e}(z)\propto
z^{-2}$.} 
\label{fig3}
\end{figure*}
We investigate the features of the $\gamma$-ray spectra produced
in such cascade process as a function of parameters describing the
acceleration of electrons in the jet. Since the Doppler factor of
the jet is unknown in the case of Cyg X-1 (the lower limit on the
jet velocity is $\sim 0.3$c -- Fender et al.~(2006) -- and the
inclination angle of the binary system is $i\approx 30^{\rm o}$),
two cases should be considered. When the Doppler factor is low, as
e.g. in the case of LS I +61 303  (Bednarek 2006b), then
we have to consider the $\gamma$-ray spectra produced not only by
electrons in the jet (directed toward the observer), but also the
$\gamma$-rays produced by electrons in the counter-jet (directed
in the opposite direction). However, for the relativistic jets with
Doppler factor $D\gg 1$, it is enough to consider only
$\gamma$-ray emission from the observer-directed 
jet since the contribution from the
counter-jet is suppressed due to relativistic effects.
Below we study the basic features of the cascade $\gamma$-ray
spectra, i.e. their dependence on the distance from the base of
the jet, on specific parameters defining the acceleration of
electrons, and their dependence on the phase of the binary system
and the spectra at specific phases. In the next section we discuss
the GeV and TeV $\gamma$-ray spectra which might be expected from
the massive binary system Cyg X-1.

\subsection{Dependence on the distance from the base of the jet}

Let us at first calculate the $\gamma$-ray spectra escaping from
the binary system toward an observer located at the inclination
angle $i = 30^{\rm o}$, These $\gamma$-rays are produced by electrons at
different parts of the jet, and their production is parameterized by the 
distance from the base
of the jet. The parameters defining the acceleration of electrons
have been chosen from the center of the  range of
values specified above (i.e. $\xi = 0.1$ and $\eta = 0.1$ - see Fig.~\ref{fig1}).
For easier interpretation of the results, we apply the
differential power law spectrum of the injected electrons of the
type $\propto E^{-2}$ (equal power per decade) up to the maximum
energies defined by $\xi$ and $\eta$. In Fig.~\ref{fig3}, we show
the cascade-produced $\gamma$-ray spectra (multiplied by the energy squared)
that are produced by electrons in the jet, the counter-jet, and
the sum of both of them, i.e. jet + counter-jet spectra. The upper
and lower panels show the spectra which are produced when the
compact object is in front of or behind the massive star,
respectively. The shape of the $\gamma$-ray spectra clearly depend
on the injection distance of electrons.  

Electrons which are
injected closer to the base of the jet are also closer to the
surface of the massive star and therefore are immersed in a stronger
soft radiation field. These cascade $\gamma$-ray spectra show a low
level of TeV emission, due to the absorption of $\gamma$-rays, and
a relatively high level of GeV emission (produced due to the
efficient cascading). Significant fluxes of TeV $\gamma$-rays can
only be produced in the upper part of the considered jet. The
cut-offs in the $\gamma$-ray spectra are determined by the maximum
energies of electrons accelerated in the jet.  
These energies increase with
the distance from the base of the jet. The $\gamma$-ray spectra
produced toward the observer in the cascade process initiated by
electrons in the counter-jet typically show higher GeV but lower
TeV emission. This is due to the fact that these $\gamma$-rays
arrive to the observer after suffering from a severe cascade process
(the angle taken from the direction between the jet-counter-jet axis 
and the direction
toward the observer is much larger). In fact, electrons in the
counter-jet barely contribute to the $\gamma$-ray spectrum
above $\sim 1$ TeV. Therefore, the sum of the jet + counter-jet
spectra (right figures) is similar to the spectra from the jet
with some increase in the GeV energy range due to the counter-jet
contribution. The counter-jet only has the effect of changing the
normalization of the total escaping spectra in the GeV energy
range.

\subsection{Gamma-ray spectra integrated along the jet}

\begin{figure}
\vskip 8.5truecm 
\includegraphics{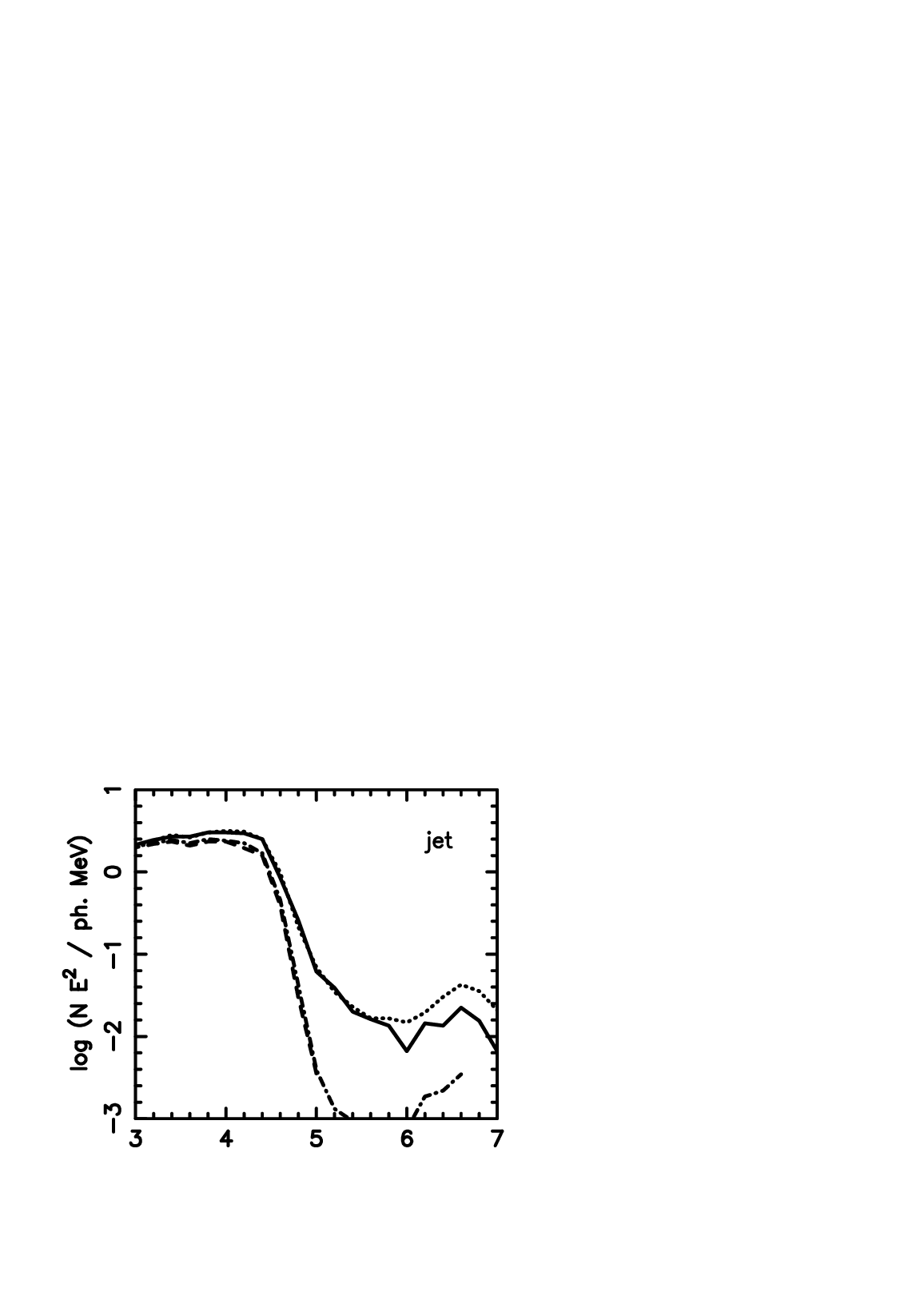} 
\includegraphics{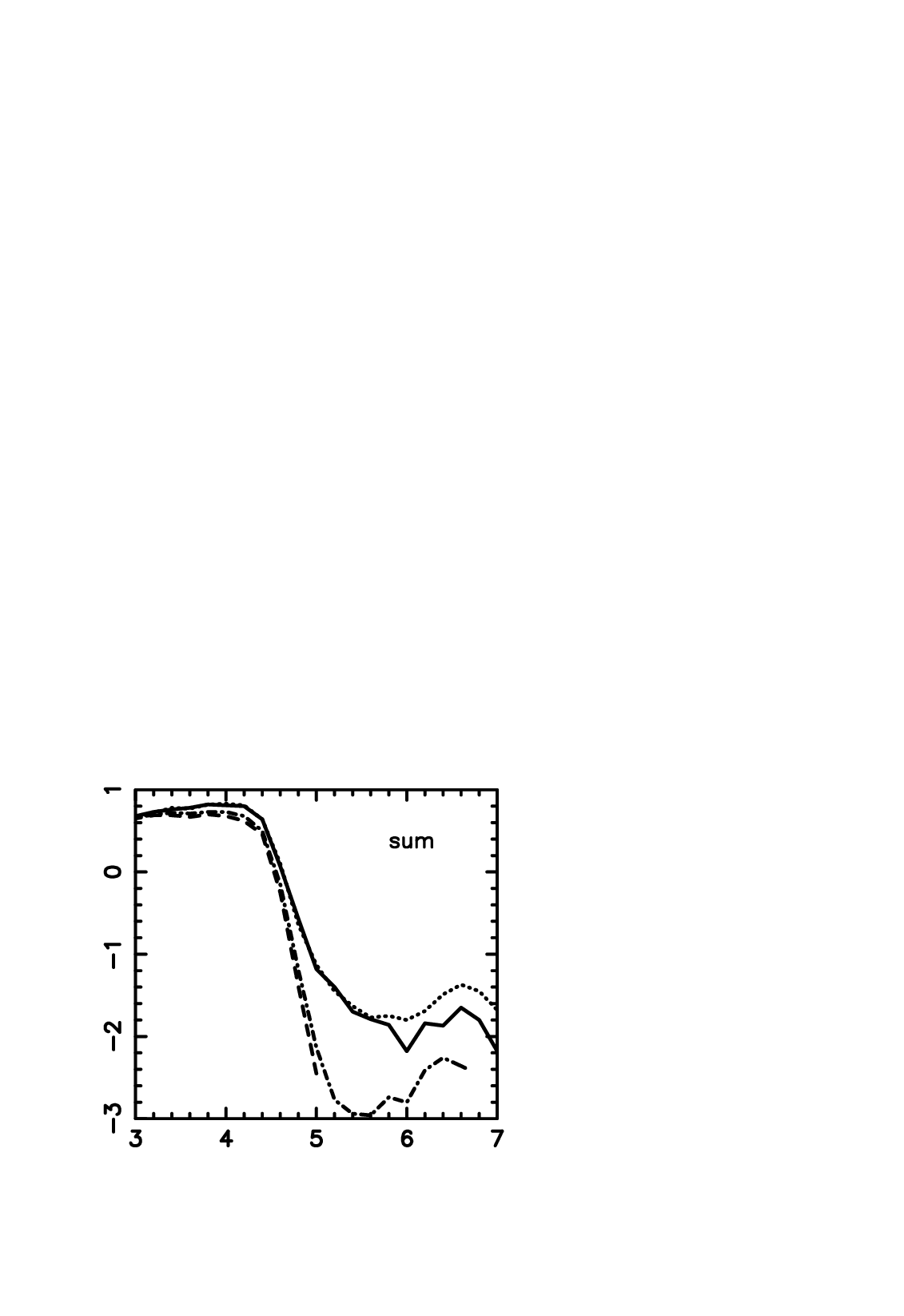}
\includegraphics{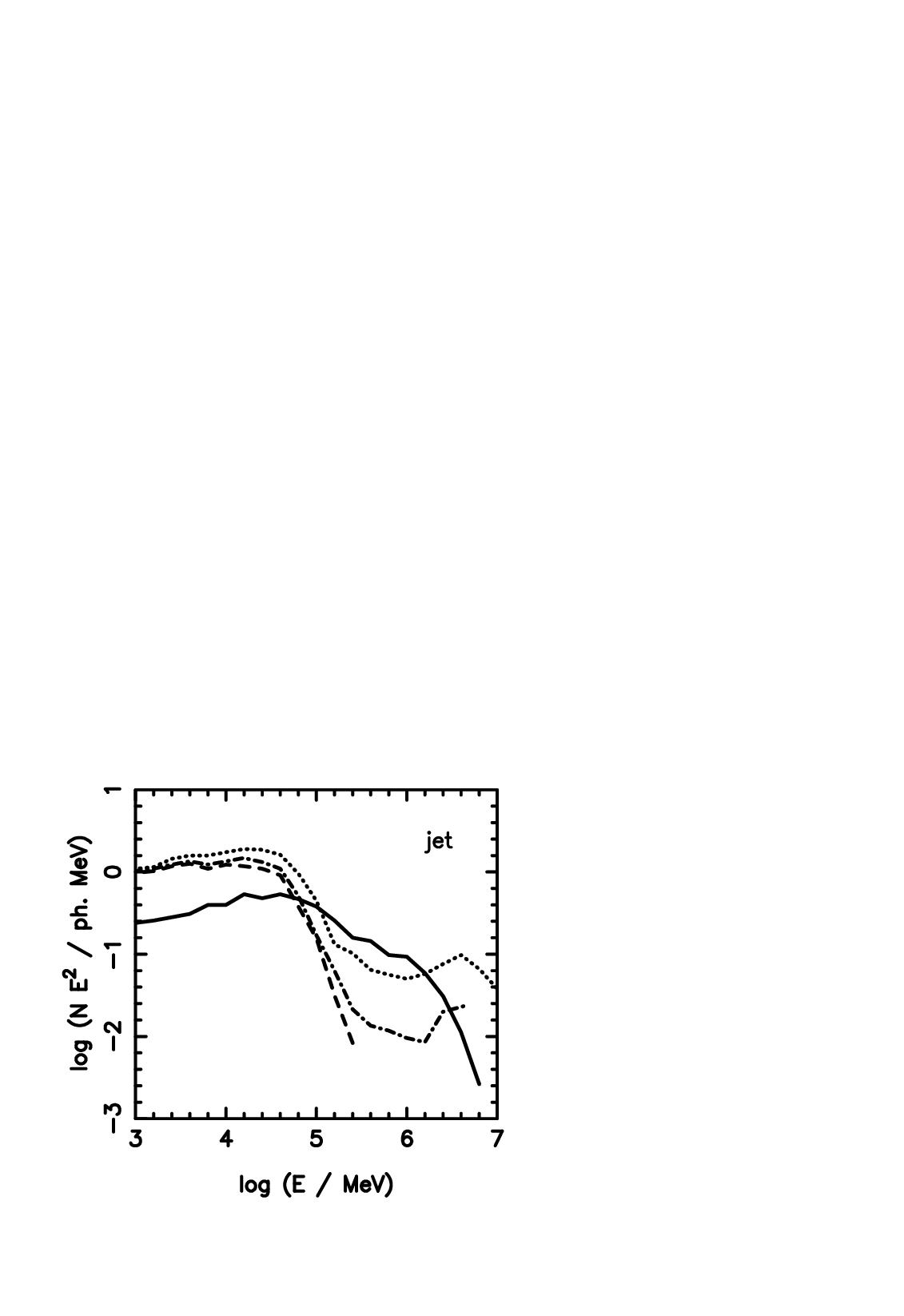} 
\includegraphics{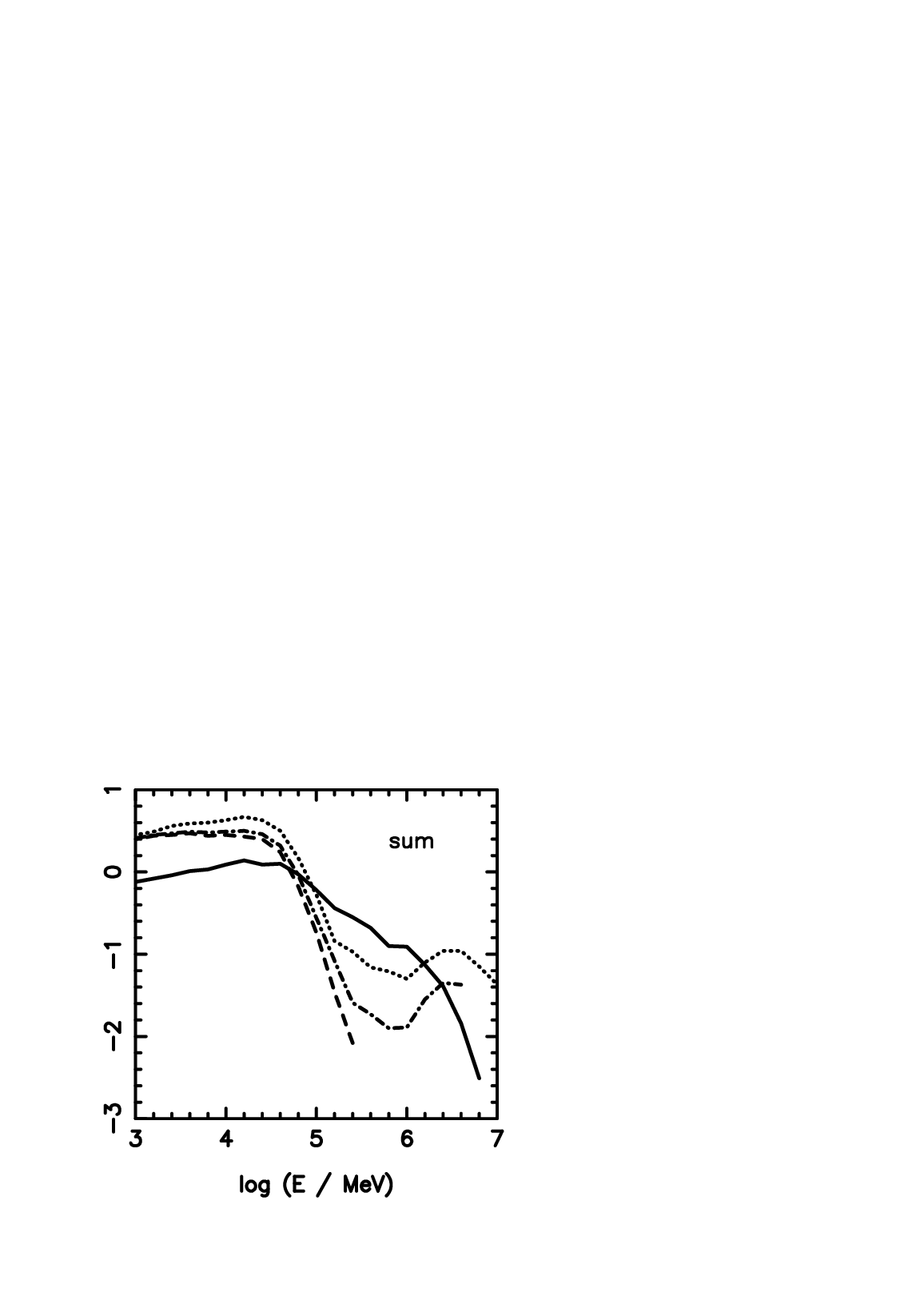} 
\caption{The cascade $\gamma$-ray spectra
(multiplied by energy square) produced by electrons in the jet
(left figures) and in the jet + counter-jet (right figures), for
the situations when the compact object is behind (upper figures)
and in front of the massive star (bottom figures). Specific
spectra have been obtained for: $\xi = 0.1$ and $\eta = 0.1$ (full
curves), $\xi = 0.01$ and $\eta = 0.1$ (dashed), $\xi = 0.1$ and
$\eta = 0.01$ (dot-dashed), $\xi = 0.3$, and $\eta = 0.1$
(dotted). They are integrated over the part of the jet extending
between z=0.1--10R$_{\star}$ measured from the base of the jet.
Electrons are injected with the power law spectrum, $\propto
E^{-2}$. The power in electron spectrum drops along the jet
according to $L_{\rm e}(z)\propto z^{-2}$.} 
\label{fig4}
\end{figure}

In Fig.~\ref{fig4}, we investigate the dependence of the cascade
$\gamma$-ray spectra escaping toward the observer as a function of the
parameters defining the acceleration model (see the profiles of
the maximum electron energies in Fig.~\ref{fig1}). The
$\gamma$-ray spectra produced in terms of this model are
integrated over the part of the jet extending between
(0.1--10)R$_\star$, since only from this limited part of the jet can
the cascade emission dominate (in the radiation of the massive star) 
and the
$\gamma$-ray emission be correlated with the orbital period of
the binary system (as recently discovered in the case of LS I +61 303, Albert et al.~2006). 

At distances closer than
0.1R$_\star$, the energy losses via synchrotron process and
scattering of the disk radiation might dominate if all electrons
can be accelerated in that region to above $\sim$1 TeV. At distances 
greater than
10R$_\star$, electrons accelerated in the jet are not able to cool
locally in the jet. They are advected at large distances along the
jet due to the motion of the plasma inside the jet, where they
cool via other processes. Note, also, that scenarios in which the
possible production of $\gamma$-rays occurs very close or far away
from the base of the jet do not provide an explanation of the correlation of the
$\gamma$-ray emission with the period of the binary system.

From a comparison of calculations performed for different
parameters, we conclude that strong TeV $\gamma$-ray emission
appears for larger values of the acceleration efficiency of
electrons in the jet (full and dotted curves in Fig.~\ref{fig4}).
As expected, the $\gamma$-ray spectra generally cut off at lower
energies for the case of less magnetized jets and lower
acceleration parameters. This is due to the saturation of electron
acceleration at lower energies by IC energy losses in these cases
(see Fig.~\ref{fig1}). The $\gamma$-ray spectra produced in the
cascade initiated by electrons, when the jet is behind the massive
star, show a sharp cut off at $\sim 100$ GeV due to the efficient
cascading for all the considered cases. However, when the compact
object is in front of the massive star, the TeV $\gamma$-ray
emission can reach the observer with a relatively flat spectrum
(spectral index $\sim$ 2.0--2.5 above a few 100 GeV) and the flux
at a level of $\sim$0.1--0.01 of the GeV emission. We conclude
that the $\gamma$-ray emission produced by electrons in the jet in
terms of the considered model has very interesting features which
should allow us to explicitly distinguish between different models
for the $\gamma$-ray production in massive binaries, e.g.
scattering of the disk radiation by electrons in the jet (as
considered in the case of active galaxies by, e.g. Dermer \&
Schlickeiser~1993) or production of $\gamma$-rays far away from
the base of the jet (e.g. Atoyan \& Aharonian~1999).

\subsection{Gamma-ray light curve and phase dependent spectra}

\begin{figure*}
\vskip 5.truecm 
\includegraphics{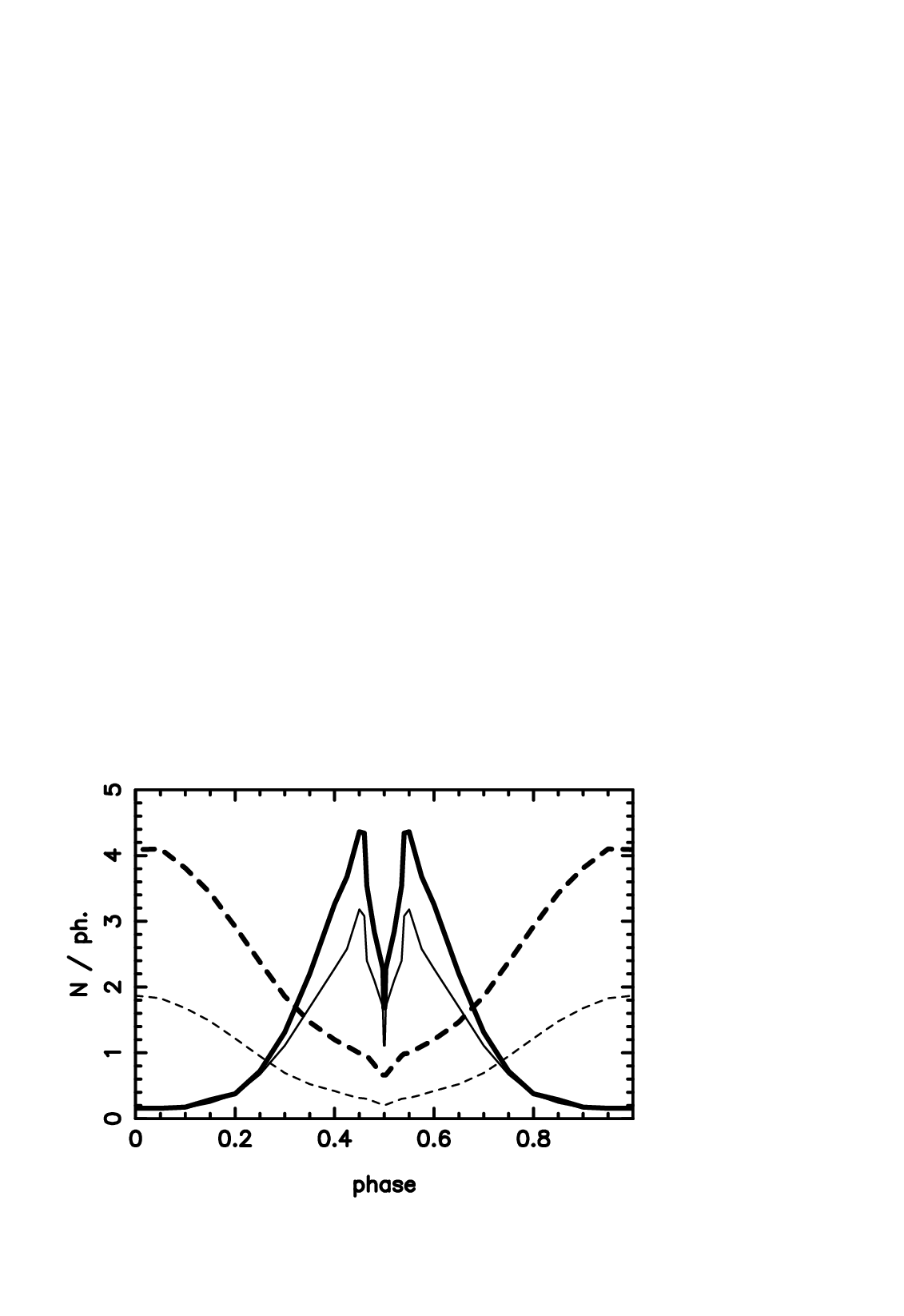} 
\includegraphics{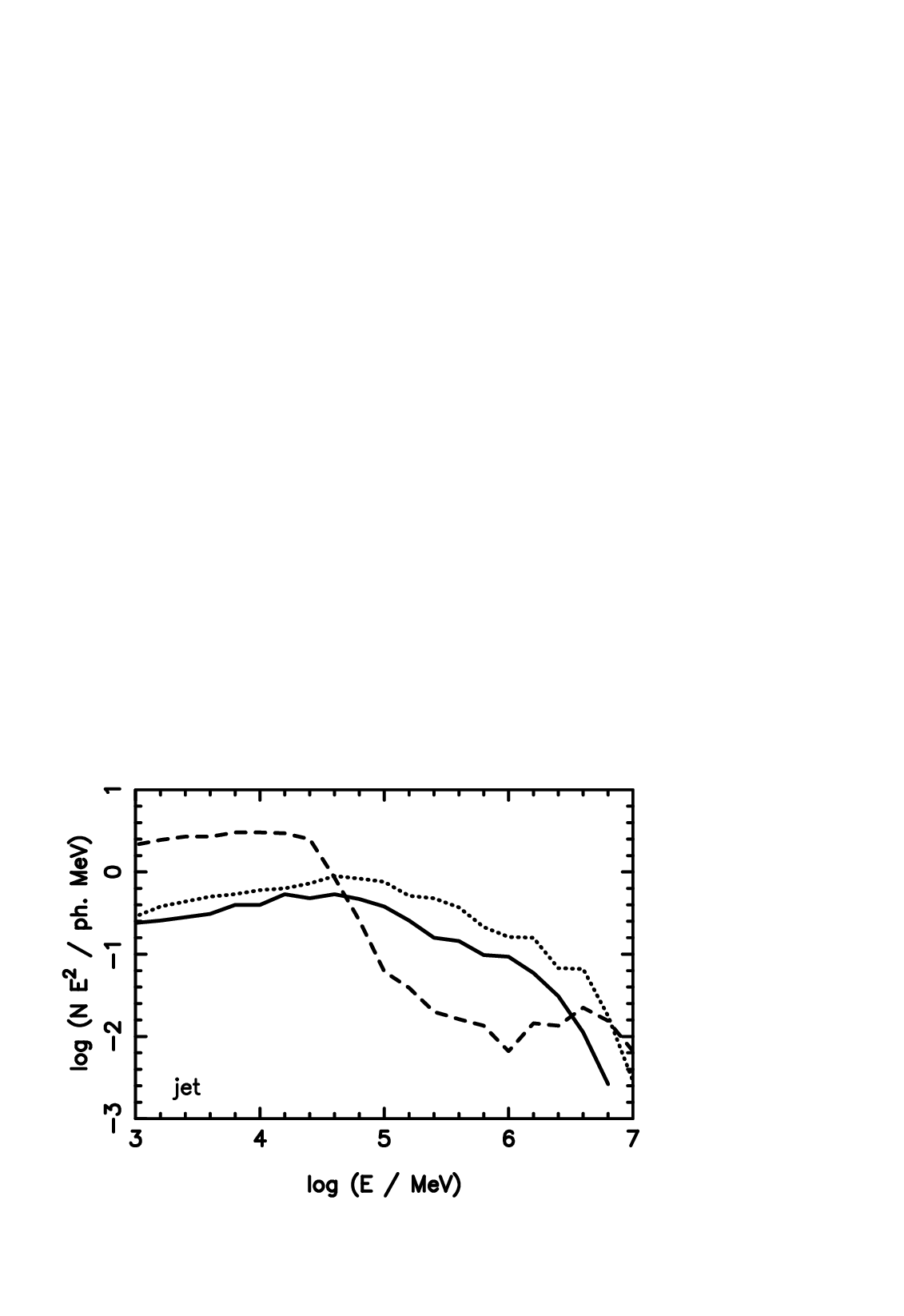}
\includegraphics{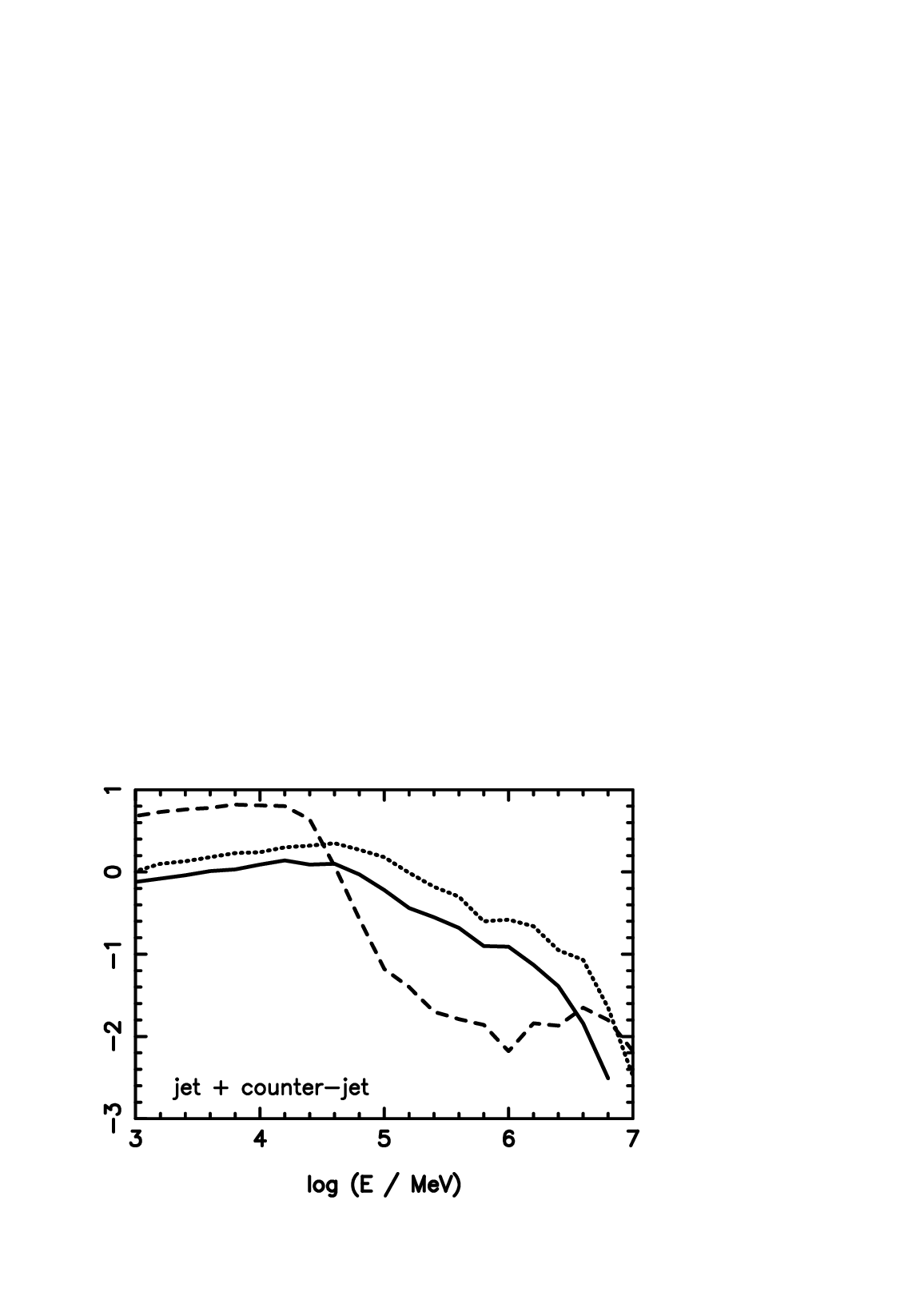} 
\caption{Gamma-ray light curves in the energy
range 1-10 GeV (dashed curves) and $>$200 GeV (full curves)
produced by electrons accelerated in the jets of Cyg X-1 for the
model defined by the parameters $\xi = 0.1$ and $\eta = 0.1$, the
power law spectrum of electrons, $\propto E^{-2}$, and the power
transferred to electrons from the acceleration mechanism which
drops along the jet according to $L_{\rm e}(z)\propto z^{-2}$.
$\gamma$-rays produced by electrons in the jet (thin curves) and
in the jet + counter-jet (thick curves). The middle and right
panels show the cascade IC $\gamma$-ray spectra escaping to the
observer from the binary system for three phases of the compact
object (measured from the location of the compact object behind
the massive star): 0. (dashed curve), 0.45 (dotted), and 0.5
(full).} 
\label{fig5}
\end{figure*}

Let us now discuss the dependence of $\gamma$-ray emission as a
function of the phase of the binary system, where the zero phase is 
taken when the 
location of the compact object is behind the massive star. Since the
orbit of the compact object is almost circular, the $\gamma$-ray
light curves should be symmetrical. We calculate the number of
$\gamma$-ray photons that escape from the binary system toward
the observer located at the inclination angle $i = 30^{\rm o}$, in
two energy ranges: the so-called GeV energy range (between 1--10
GeV available for the past EGRET telescope and the future AGILE
and GLAST telescopes), and the TeV energy range ($>$200 GeV)
available for the present and future Cherenkov telescopes on the
northern hemisphere such as the MAGIC and VERITAS telescopes. As
an example, we concentrate on the model defined by $\xi = 0.1$ and
$\eta = 0.1$ (the parameters from the center of considered range),
and the electron spectrum described by $\propto E^{-2}$, and its
power dropping along the jet according to $L_{\rm e}(z)\propto
z^{-2}$. The $\gamma$-ray light curves (number of photons arriving
to the observer in the GeV and TeV energy ranges) are shown in
Fig.~\ref{fig5} for the case of $\gamma$-ray emission only from
the jet (thin curves) and the sum of the emission from the jet +
counter-jet (thick curves). The TeV $\gamma$-ray emission can
change in this case by more than a factor of 20, showing the
minimum when the compact object is behind the massive star and a
sharp maximum for the phases at $\sim$0.05 before and after the
situation when the compact object is in front of the massive star.

Exactly in front of the massive star, the TeV $\gamma$-ray flux
becomes a factor of $\sim$2 lower. Due to the lowest
optical depths for $\gamma$-rays and leptons
production of $\gamma$-rays is less favourite in such a geometry
than for surrounding phases. The contribution of the
counter-jet to the maximum in the TeV $\gamma$-ray light curve is
at the level of $\sim 30\%$ of the jet emission (compare the thin
and thick full curves). On the other hand, the GeV $\gamma$-ray
light curve behaves differently. It has the maximum when the
compact object is behind the massive star and minimum exactly when
the compact object is in front of the massive star. The maximum is
due to the fact that the $\gamma$-ray spectrum is formed in the very
efficient cascade process when the compact object and the
observer are on the opposite sites of the massive star. The
contributions of the jet and the counter-jet to the GeV
$\gamma$-ray light curve are comparable. It is remarkable that there 
is a very
interesting anticorrelation between the $\gamma$-ray light curves
in the GeV and TeV energy ranges. This is the unique feature which
can serve as an important diagnostic of this model for the high
energy processes occurring in the $\gamma$-ray emitting massive
binary systems.

Fig.~\ref{fig5} also shows the
$\gamma$-ray spectra emerging toward the observer for some
characteristic phases of the binary system for the case of the
acceleration of electrons in the jet and the sum for the jet +
counter-jet emission. Note the dependence between the
spectral shape and the flux of the TeV $\gamma$-ray emission. It
is predicted that the lower level of $\gamma$-ray flux should be
connected with a flattening of the spectrum above $\sim$200 GeV and
vice versa. The contribution of $\gamma$-ray emission from the
counter-jet has an effect of increasing the GeV $\gamma$-ray flux
with no
effect on the TeV flux. However, due to the higher GeV
emission, the relative power in the GeV and TeV energy ranges
changes. This has consequences for predicting the TeV flux from
the normalization to the GeV (EGRET) observations.

\section{The multifrequency spectrum of Cyg X-1}

The $\gamma$-ray spectrum of Cyg X-1 is in fact very poorly
constrained. The spectrum in the soft X-ray state shows the power
law tail with the differential spectral index -2.6 extending up to
$\sim$10 MeV, without evidence of a cut off (McConnell et
al.~2000, 2002). However, the jet appears only in the low-hard state, so this emission
can be only considered as an upper limit on the MeV $\gamma$-ray flux related to the jet.
The upper limit at higher energies ($>100$ MeV) has been derived (see McConnell et al.~2000), 
based on the 3rd EGRET catalog (Hartman et al.~1999).

\begin{figure*}
\vskip 9.truecm 
\includegraphics{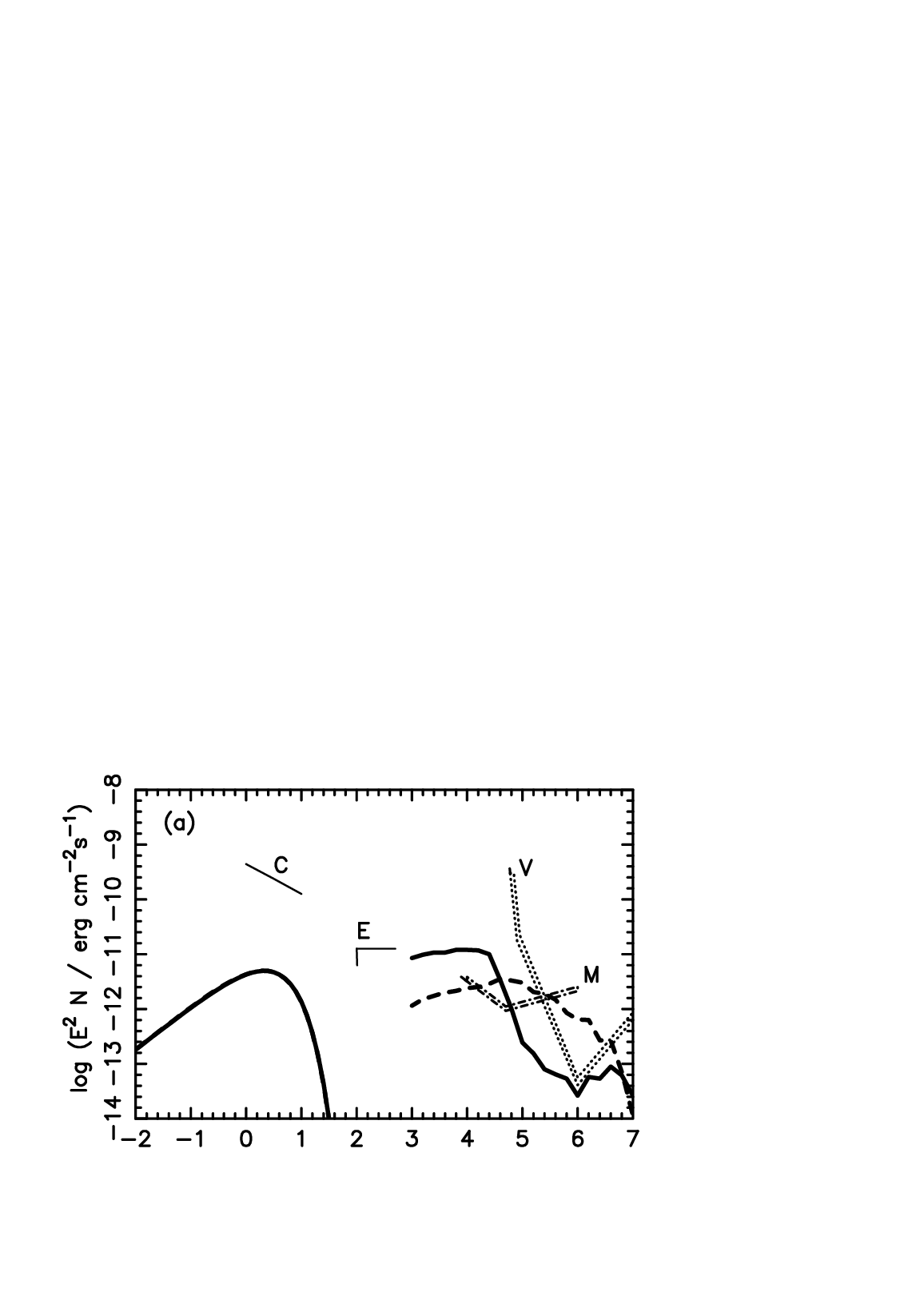} 
\includegraphics{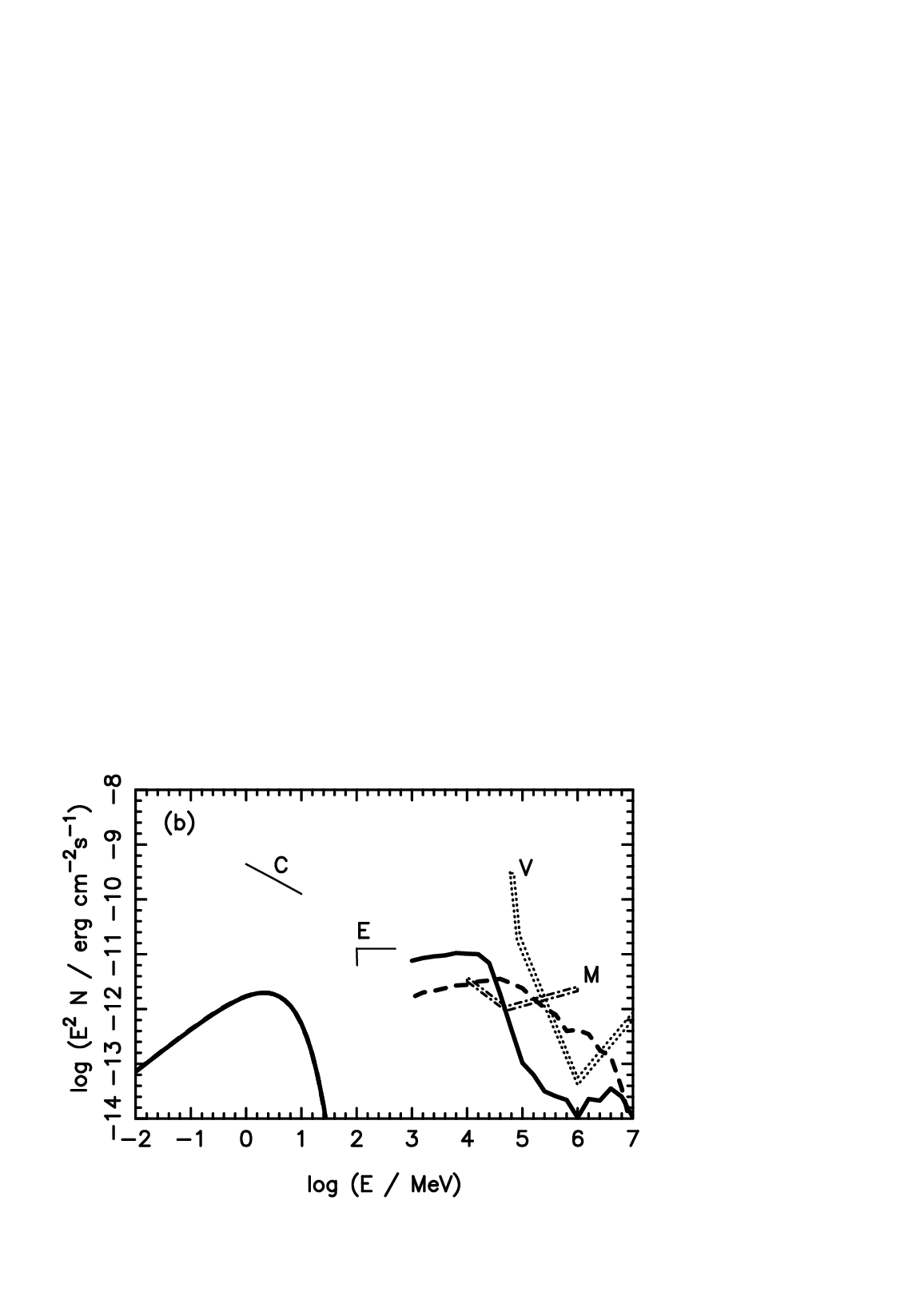}
\includegraphics{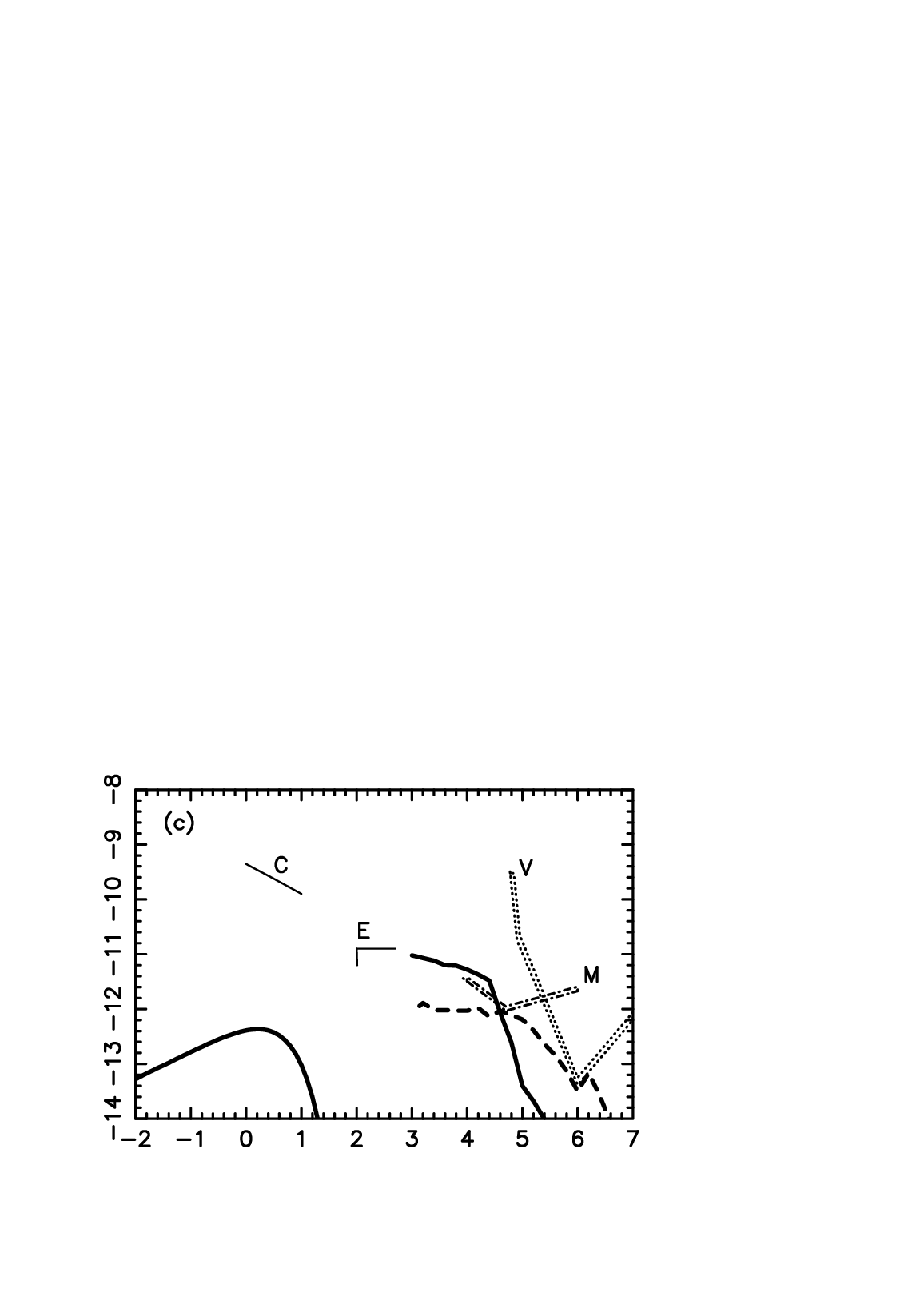} 
\includegraphics{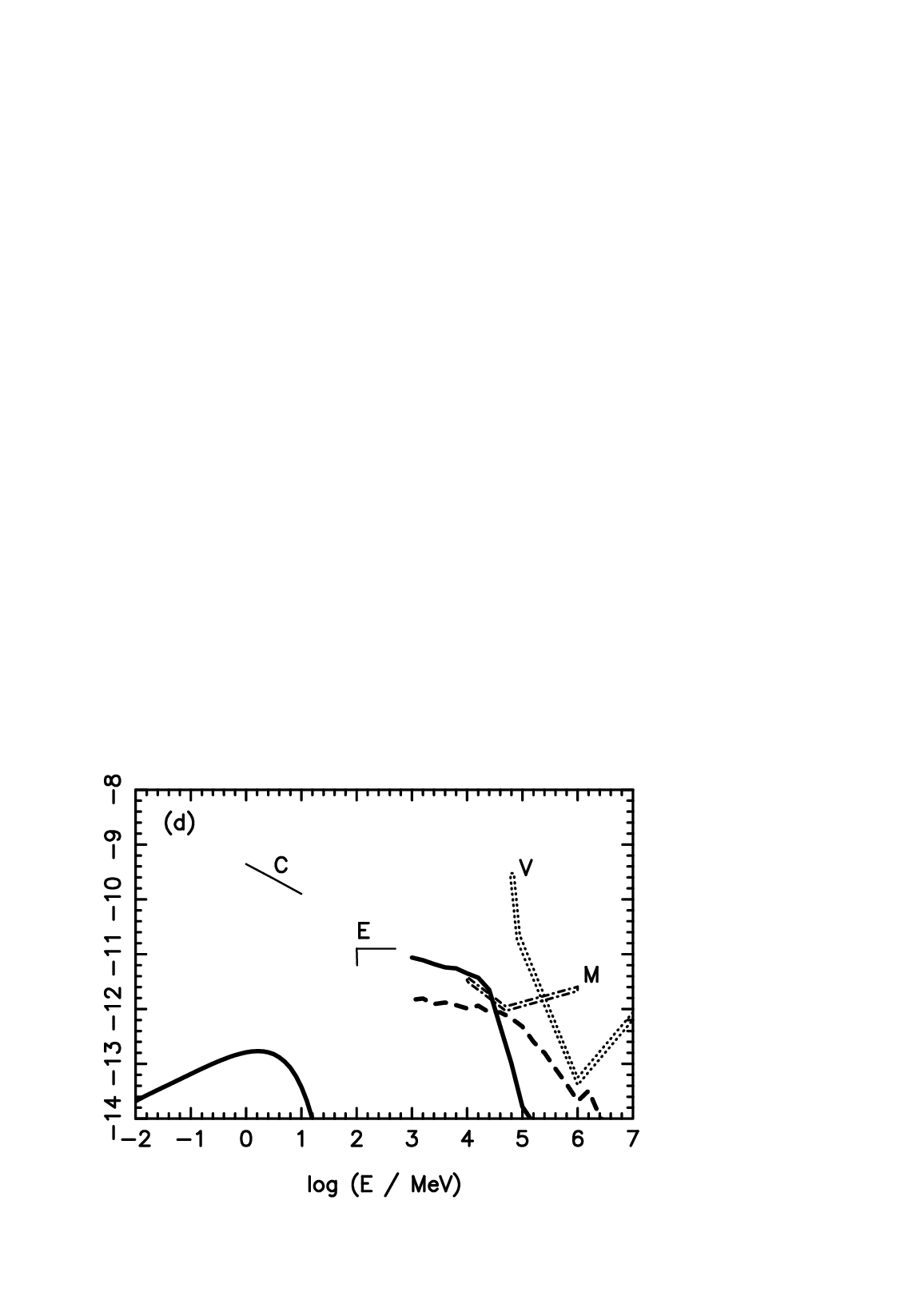}
\includegraphics{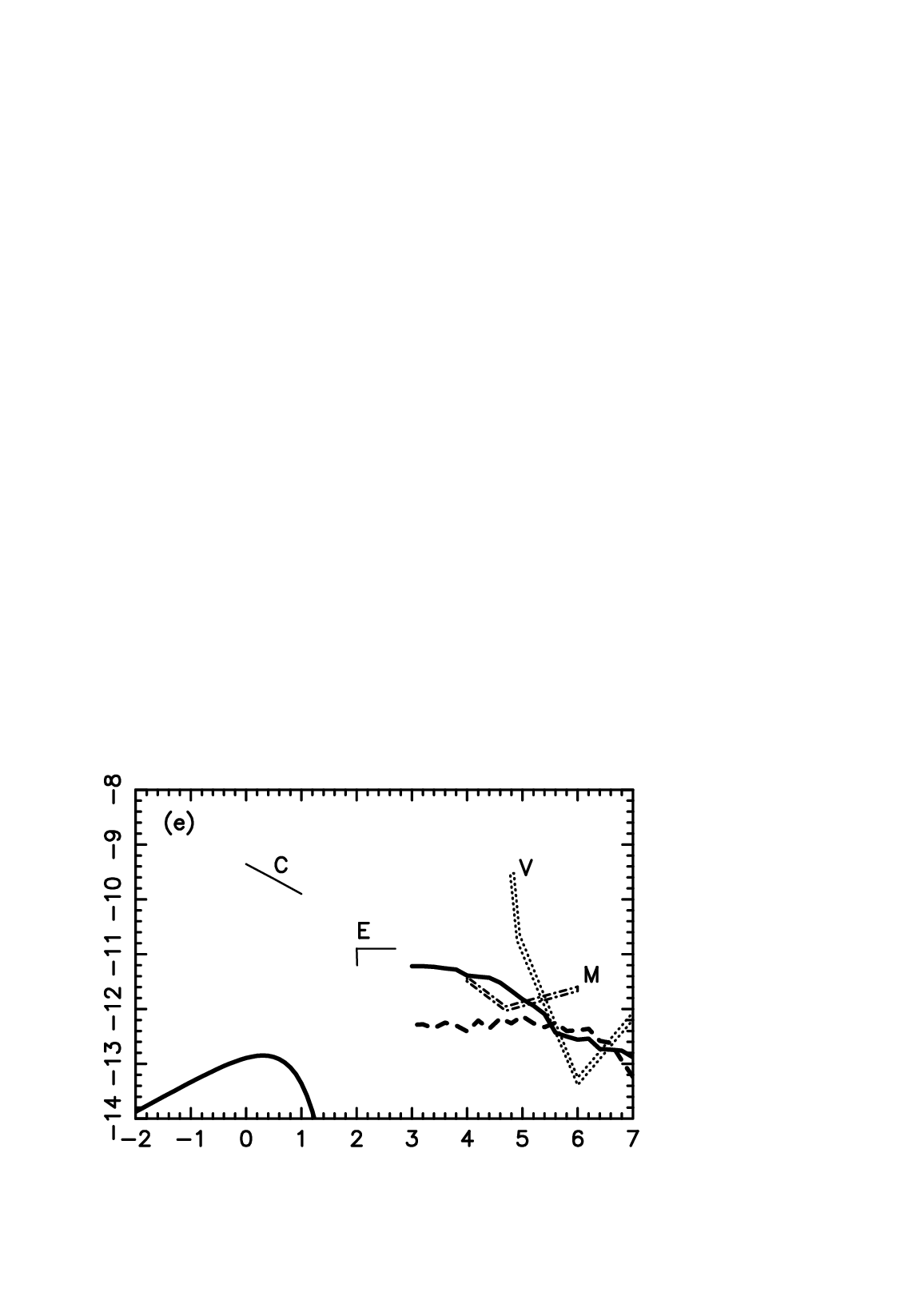} 
\includegraphics{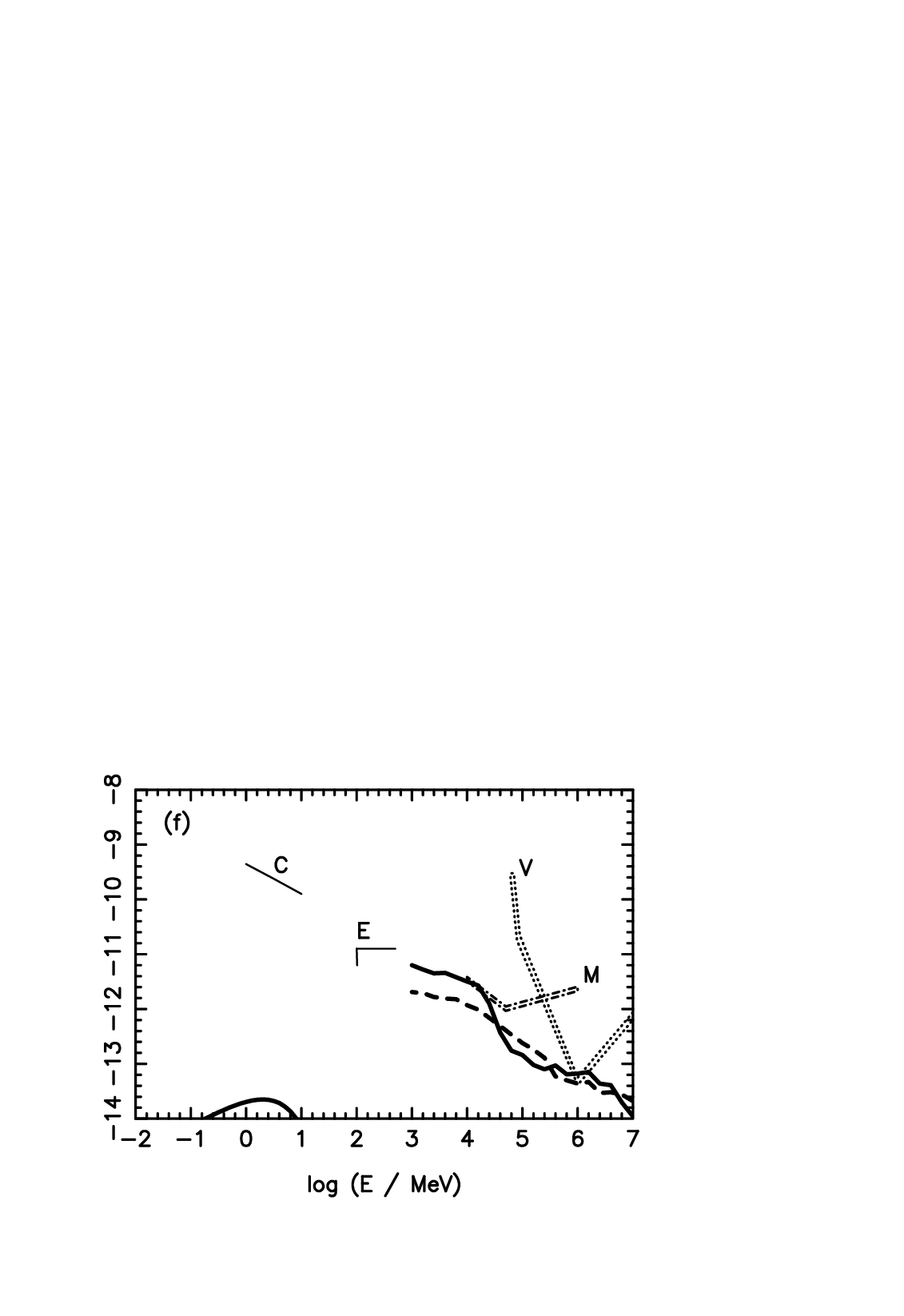} 
\caption{The
multiwavelength spectrum of Cyg X-1 (differential spectrum
multiplied by energy squared) is compared with the calculated
cascade $\gamma$-ray spectra for three  acceleration models of
electrons in the jet:  (1) differential spectral index of
electrons $\propto E^{-\alpha}$ and $\alpha = 2$, and the power
transfered to electrons changes with distance along the jet
according to $L_{\rm e}(z)\propto z^{-2}$ (model A); (2) $\alpha =
2.6$ and  $L_{\rm e}(z)\propto z^{-2}$ (model B), (3) $\alpha =
2.6$ and  $L_{\rm e}(z) = const$ (model C). The cut-off in the
electron spectrum is determined by $\xi = 0.1$, $\eta = 0.1$. The
upper and bottom panels show the spectra produced in the jet and
the jet + counter-jet, respectively. The $\gamma$-ray spectra
above are shown for the minimum flux above 100 GeV, i.e. at  the
phase 0. counted from the location of the compact object behind
the massive star (full curves), and for the maximum flux at the
phase $\sim$0.45 and 0.55 (dashed curves). The synchrotron
spectrum produced by electrons in the jet are shown by the full
curves below MeV energies.  The spectrum from Cyg X-1 reported by
the COMPTEL telescope and the upper limit obtained based on the
EGRET data (McConnell et al.~2000, 2002) are shown by thin full
curves and marked by C and E, respectively. The sensitivities of
the MAGIC and VERITAS telescopes are shown by double dot-dashed
and dotted curves and marked by M and V, respectively (see e.g.
Lorenz~2000).} \label{fig6}
\end{figure*}

However, in the context of the recent discoveries of the TeV
emission from massive binary systems  (i.e. LS 5039 and LS I +61 303),
which have similar orbital parameters to Cyg
X-1 (i.e., a massive star and separation of the components), it is 
reasonable to
predict the level and spectral features of the possible high
energy $\gamma$-ray emission also from this object. Based on the
IC $e^\pm$ pair cascade model (Bednarek~2006b), considered here
for Cyg X-1, we were able to predict the phase dependent TeV
emission from LS I +61 303 (Albert et al.~2006). In the
case of Cyg X-1, we normalize the cascade $\gamma$-ray spectra
expected from this object to the upper limit derived from the
EGRET data.  Note that, the MeV $\gamma$-ray emission from Cyg X-1 is 
$\sim 1\%$ of its total bolometric power (McConnell et al.~2002), and the EGRET
upper limit in the GeV energies is again on the level of $10\%$ of the MeV emission.
On the other hand, in the cascade model which we consider almost all energy from 
relativistic electrons in the jet is transfered to  gamma-rays (and also secondary pairs), 
see  Bednarek~(2006b).
Since the jet power is expected to be comparable to the observed luminosity of the Cyg X-1,
we conclude that the considered model is consistent if only a small part of jet energy is
converted  into  relativistic electrons (i.e. $\sim 0.1\%$). 
The applied normalization to the EGRET upper limit allows us to estimate the level of
the TeV $\gamma$-ray flux and spectrum from Cyg X-1. In
Figs.~\ref{fig6}, we show the comparison of the results of
calculations with the observational constraints at lower energies.
The upper and the lower panels  we show the cascade $\gamma$-ray
spectra escaping toward the observer located at the inclination
angle $i = 30^{\rm o}$ for the case of their production in the jet
and the jet+counter-jet (the sum of both), respectively. Since we
have no detailed information about the spectral shape in the GeV
energy range from the EGRET observations (but only the upper
limit), electrons injected into the jet with different spectra and
injection efficiencies along the jet have to be considered. At
first we consider the model with very flat electron spectrum
(differential spectral index equal to -2) and the power
transferred to electrons which changes with distance along the jet
according to $L_{\rm e}(z)\propto z^{-2}$ (model A). In fact, such
flat electron spectrum seems to be not very realistic if we
compare the spectral features of the source LS I +61 303 
(the object of our previous modeling) and the spectrum of Cyg
X-1. The electron spectrum with the index 2 postulates the
appearance of a completely new component in the high energy
spectrum of Cyg X-1 which does not lay on the extrapolation from
the MeV energy range. The comparison of the calculated cascade
spectra with the available data and the sensitivities of the
present MAGIC telescope (M) and the future VERITAS telescope (V)
is shown in Fig.~\ref{fig6}a,b. The dashed and full curves show
the expected spectra when the compact object is in front of and
behind the massive star  HDE~226868. Clearly, for such model
parameters, Cyg X-1 can be potentially detected at energies above
$\sim$100 GeV by the MAGIC telescope and above a few 100 GeV by
the VERITAS telescope when the compact object is in front of the
massive star. We also show the synchrotron spectrum which is
produced by electrons accelerated in the jets assuming that the
Doppler boosting is not important (bumps at lower energies marked
by full curves). These spectra are at a much lower level than the
observations of Cyg X-1 in the MeV energy range and cannot be
responsible for this emission. Note, however, that if  significant
Doppler boosting is discovered in Cyg X-1 (the speed of the jet
comparable to the velocity of light), the flux in the calculated
synchrotron spectrum should go up with respect to the $\gamma$-ray
spectra. This is due to the fact that the cascade $\gamma$-ray
spectra are mainly formed outside the jet, i.e. inside the whole
volume of the binary system where the plasma move with the wind
velocity of the massive star. In such a case, the Doppler boosting
can be only important for the synchrotron radiation but is clearly
unimportant for cascade $\gamma$-rays.
This MeV $\gamma$-ray emission can also originate in another 
scenario, e.g. in the hot corona of the accretion disk (e.g. 
Zdziarski \& Gierli\'nski~2004, Chakrabarti \& Mandal~2006) 
or in the inner part
of the jet close to its base (a scenario not considered in this paper) 
as a result of Comptonization 
of the disk radiation and possible IC cascading in the disk radiation, 
e.g. Aharonian et al.~1985, Coppi, Blandford \& Rees~1993).

The calculations with the more realistic model parameters (see Table.~2) are
compared with the Cyg X-1 data and sensitivities of the Cherenkov
telescopes and are shown in Figs.~\ref{fig6}. We show in
Figs.~\ref{fig6}c the $\gamma$-ray spectra from the jet and in
Figs.~\ref{fig6}d the jet + counter-jet, which are calculated for
electrons with the differential spectral index $\propto E^{-2.6}$,
and the power in electrons dropping along the jet according to
$L_{\rm e}(z)\propto z^{-2}$ (model B). For these parameters,
there is no chance for detection of TeV $\gamma$-ray emission
neither by the MAGIC telescope above $\sim$100 GeV nor by the
Veritas telescope at $\sim$1 TeV. Finally in Figs.~\ref{fig6}ef,
the $\gamma$-ray spectra are shown for electrons injected with the
steep spectrum ($\propto E^{-2.6}$) but with flat profile for the
power transferred from the acceleration mechanism to electrons,
$L_{\rm e}(z)=const$ (model C). For this model, a large power is
transferred to electrons at larger distances from the base of the
jet where the acceleration of electrons is saturated at larger
energies. Then, the cascade $\gamma$-ray spectra at TeV energies
are relatively flat, in spite of the strong $\gamma$-ray
absorption effects. The $\gamma$-ray spectra produced from the
jet, after normalization to the EGRET upper limit in the GeV
energy range, have chance to be detected by the VERITAS telescope
at TeV energies but not by the MAGIC telescope above $\sim$100 GeV
(Figure 6e). Note that the unboosted synchrotron spectra
calculated in model B and C are at the lower level than in the
model A due to different normalization in the GeV energy range.

%
\begin{table}
\caption{Parameetrs of the models considered in Fig.~\ref{fig6}. }             
\label{tab2}      
\begin{tabular}{c c c c}     
\hline\hline       
Model &  A & B & C\\ 
\hline                    
differential spectrum                  &                &                   &                    \\ 
of electrons:  $N(E)\propto$          & $E^{-2}$ & $E^{-2.6}$ & $E^{-2.6}$  \\                                     \hline
injection number  of electrons      &   &   &                                               \\ 
along the jet:   $L_{\rm e}(z)\propto$       &  $z^{-2}$ & $z^{-2}$ &  $const.$       \\ 
\hline                  
\end{tabular}
\end{table}
\section{Conclusions}

We constrain the parameters of the IC $e^\pm$ pair cascade model
for which we predict observable $\gamma$-ray fluxes from Cyg X-1
at energies $>$100 GeV by the MAGIC and/or VERITAS telescopes. The
main conclusions from the analysis of this model for the
$\gamma$-ray production in microquasars are reported in the
following.

The detection of the TeV $\gamma$-ray emission from Cyg X-1 by
Cherenkov telescopes can occur only under very special conditions
for the acceleration of electrons in the jet as described in the model 
herein considered, and for an acceleration efficiency parameter $\xi$,
the magnetic field strength parameter $\eta$, a flat spectrum of
electrons, and a flat profile of the conversion of energy from the
acceleration mechanism to electrons along the jet. These
conditions are not very likely and do not fit nicely to
$\gamma$-ray observations of Cyg X-1 at the MeV-GeV energy range.

Provided that such strict conditions for the acceleration
of electrons are met (e.g. as in the considered model A), we have
found that in the case of compact massive binary systems with
circular orbits of the Cyg X-1 type, the highest chances for
detection of TeV $\gamma$-ray emission by the Cherenkov telescopes
are for phases $\sim 0.45$ and $\sim 0.55$, counted from the
position of the compact object behind the massive star. The flux
expected at the moment when the compact object is exactly in front
of the massive star is a factor of $\sim$2 lower.
The GeV and TeV $\gamma$-ray fluxes should be anticorrelated.
The largest GeV flux is expected when the compact object is behind
the massive star. The GeV $\gamma$-ray flux should vary by a factor of $\sim$5
and the TeV flux by a factor of $\sim$20 with the orbital period
of Cyg X-1/HDE226868 system.

We want to remind that the above conclusions concerning possible
detection of the very high energy $\gamma$-ray emission from Cyg
X-1 are based on the assumption that the upper limit derived from
the EGRET observations provides correct constraints on the
multifrequency spectrum of Cyg X-1, independent of the activity
state of this binary system. In fact, this upper limit is
consistent with the extrapolation of the $\gamma$-ray spectrum
reported by the COMPTEL telescope from the Cyg X-1 region during
the hard X-ray state of this source. Such flat MeV spectrum is not
observed from Cyg X-1 during its other activity states (e.g.
McConnell et al.~2000,2002).

\begin{acknowledgements}
We thank the referee for valuable comments and James H. Beall for reading
the manuscript and comments.
This work is supported by the Polish MNiI grant No. 1P03D01028.
\end{acknowledgements}

\end{document}